\documentclass[a4paper,11pt]{article}
\usepackage[dvips]{graphicx}
\usepackage{amsmath,amssymb}
\usepackage[T1]{fontenc}
\usepackage{xspace}
\usepackage{rotating}
\usepackage{mathrsfs}
\begin{document}
%
%
\renewcommand{\floatpagefraction}{0.7}
\renewcommand{\textfraction}{0.1}
\renewcommand{\bottomfraction}{0.6}
\newcommand{\Szero}{\ensuremath{\mathrm{S}^0}\xspace}
\newcommand{\mS}{\ensuremath{m_{\Szero}}\xspace}
\newcommand{\HSM}{\ensuremath{\mathrm{H}^0_{\mathrm{SM}}}\xspace}
\newcommand{\OPAL}{{\small OPAL}\xspace}
\newcommand{\LEP}{{\small LEP}\xspace}
\newcommand{\Zzero}{\ensuremath{\mathrm{Z}^{0}}\xspace}
\newcommand{\hzero}{\ensuremath{\mathrm{h}^{0}}\xspace}
\newcommand{\Hzero}{\ensuremath{\mathrm{H}^{0}}\xspace}
\newcommand{\dEdx}{\ensuremath{\mathrm{d}E/\mathrm{d}x}\xspace}
\newcommand{\gce}{{\small GCE}\xspace}
\newcommand{\sm}{{\small SM}\xspace}
\newcommand{\SM}{{Standard Model}\xspace}
\newcommand{\MSSM}{{\small MSSM}\xspace}
\newcommand{\mc}{Monte Carlo\xspace}
\newcommand{\MC}{Monte Carlo\xspace}
\newcommand{\klein}[1]{{\small #1}\xspace}
\newcommand{\hl}[1]{{\itshape #1}}
\newcommand{\degree}[1]{\ensuremath{\mathrm{#1}^\circ}}
\newcommand{\pa}{\ensuremath{\phi_a}\xspace}
\newcommand{\wa}{\ensuremath{\alpha}\xspace}
\newcommand{\aiso}{\ensuremath{\alpha_{\mathrm{iso}}}\xspace}
\newcommand{\minv}{\ensuremath{m_{\mathrm{inv}}}\xspace}
\newcommand{\tpmiss}{\ensuremath{\theta(\vec{p}_{\mathrm{miss}})}\xspace}
\newcommand{\pmiss}{\ensuremath{p_{\mathrm{miss}}}\xspace}
\newcommand{\nn}{\ensuremath{\nu\overline{\nu}}\xspace}
\newcommand{\mm}{\ensuremath{\mu^+\mu^-}\xspace}
\newcommand{\lplm}{\ensuremath{\mathrm{l}^+\mathrm{l}^-}\xspace}
\newcommand{\ee}{\ensuremath{\mathrm{e}^+\mathrm{e}^-}\xspace}
\newcommand{\bbar}{\ensuremath{\mathrm{b\overline{b}}}\xspace}
\newcommand{\keV}{\ensuremath{\mbox{keV}}\xspace}
\newcommand{\MeV}{\ensuremath{\mbox{MeV}}\xspace}
\newcommand{\GeV}{\ensuremath{\mbox{GeV}}\xspace}
\newcommand{\gev}{\ensuremath{\mbox{GeV}}\xspace}
\newcommand{\Evec}{\ensuremath{\vec{E}}\xspace}
\newcommand{\tbc}{(\emph{\ldots to be completed \ldots})\xspace}
\newcommand{\mrec}{\ensuremath{m_{\mathrm{r}}}\xspace}
\newcommand{\sq}{\ensuremath{k}\xspace}
\newcommand{\Nnf}{\ensuremath{\mathrm{N^{95}}}\xspace}
\newcommand{\sqnf}{\ensuremath{\mathrm{k^{95}}}\xspace}
\newcommand{\Nsm}{\ensuremath{\mathrm{N_{SM}}}\xspace}
\newcommand{\THDM}{{\small 2HDM}\xspace}
\newcommand{\SigmaZH}{\ensuremath{\sigma_{\mathrm{ZH^{SM}}}}\xspace}
\newcommand{\mH}{\ensuremath{m_{\mathrm{H}}^{\mathrm{SM}}}\xspace}
\newcommand{\mHsm}{\ensuremath{m_{\mathrm{H}}}\xspace}
\newcommand{\mh}{\ensuremath{m_{\mathrm{h^0}}}\xspace}
\newcommand{\mhwo}{\ensuremath{m}\xspace}
\newcommand{\mhi}{\ensuremath{m_{\mathrm{h^0_i}}}\xspace}
\newcommand{\hi}{\ensuremath{\mathrm{h^0_i}}\xspace}
\newcommand{\Ki}{\ensuremath{K_i}\xspace}
\newcommand{\SigmaZh}{\ensuremath{\sigma_{\mathrm{Zh}}}\xspace}
\newcommand{\mA}{\ensuremath{m_{\mathrm{A}}}\xspace}
\newcommand{\mB}{\ensuremath{m_{\mathrm{B}}}\xspace}
\newcommand{\mC}{\ensuremath{m_{\mathrm{C}}}\xspace}
\newcommand{\intKdm}{\ensuremath{\int \tilde{K}\mathrm{d}m}\xspace}
\newcommand{\Ktilde}{\ensuremath{\tilde{K}}\xspace}
\newcommand{\err}[3]{\,\ensuremath{\mathrm{#1}\pm\mathrm{#2\,(stat.)}\pm\mathrm{#3\,(syst.)}}\xspace}
\newcommand{\erro}[2]{\,\ensuremath{\pm\mathrm{#1}\pm\mathrm{#2}}\xspace}
\newcommand{\intk}{\ensuremath{\int_{\mA}^{\mB} \tilde{K}\,\mathrm{d}m}\xspace}
\newcommand{\mathd}{\ensuremath{\mathrm{d}}}
\newcommand{\deltam}{\ensuremath{\Delta m}}
\newcommand{\rb}[1]{\raisebox{-2ex}{#1}}
\newcommand{\pb}{\ensuremath{\mathrm{pb}^{-1}}}
\newcommand{\Ztoee}{\ensuremath{\Zzero\to\ee}}
\newcommand{\Ztomm}{\ensuremath{\Zzero\to\mm}}
\newcommand{\vnr}[1]{\vphantom{\rule{0mm}{#1}}\xspace}
\newcommand{\rr}{\raggedright\small}
\newcommand{\prelim}{\large\textbf{Preliminary}\xspace}
\newcommand{\G}{\mbox{$\mathrm{GeV}$}}
\newcommand{\sqrts}{\mbox{$\sqrt {s}$}}
\newcommand{\mZ}{$m_{\mathrm{Z}}$\xspace}
\newcommand{\bb}{\mbox{$\mathrm{b}\bar{\mathrm{b}}$}}
\newcommand{\etal}{\mbox{\it et al.}}
%
%
\topsep0pt plus 1pt
\begin{titlepage}
\begin{center}
  {\large  EUROPEAN ORGANIZATION FOR NUCLEAR RESEARCH}
\end{center}
\bigskip
\begin{flushright}
  CERN-EP-2002-032   \\ 14. May 2002
\end{flushright}
\bigskip\bigskip\bigskip\bigskip\bigskip
\begin{center}
  {\huge\bfseries
    Decay-mode independent searches for new\\[1.2ex]
    scalar bosons with the {OPAL detector} at LEP
  }
\end{center}
\bigskip\bigskip
\begin{center}
  {\LARGE The OPAL Collaboration}
\end{center}
\bigskip\bigskip\bigskip
\begin{center}
  {\large  Abstract}
\end{center}
{ \noindent This paper describes topological searches for neutral
  scalar bosons \Szero produced in association with a \Zzero boson
  via the Bjorken process $\ee\to\Szero{}\Zzero$ at centre-of-mass
  energies of 91~\GeV and 183--209~\GeV.  These searches are based
  on studies of the recoil mass spectrum of $\Zzero\to\ee$ and
  $\mu^+ \mu^-$ events and on a search for $\Szero\Zzero$ with
  $\Zzero \to \nu\bar{\nu}$ and $\Szero \to \ee$ or photons.  They
  cover the decays of the \Szero into an arbitrary combination of
  hadrons, leptons, photons and invisible particles as well as the
  possibility that it might be stable.
  
  No indication for a signal is found in the data and upper limits
  on the cross section of the Bjorken process are calculated.
  Cross-section limits are given in terms of a scale factor \sq
  with respect to the \SM cross section for the Higgs-strahlung
  process $\ee\to\HSM\Zzero$.
  
  These results can be interpreted in general scenarios
  independently of the decay modes of the \Szero. The examples
  considered here are the production of a single new scalar particle
  with a decay width smaller than the detector mass resolution, and
  for the first time, two scenarios with continuous mass
  distributions, due to a single very broad state or several states
  close in mass.
  }
\bigskip\bigskip\bigskip\bigskip\bigskip\bigskip
\begin{center}
  {\large (Submitted to Eur. Phys. J.)}
\end{center}
\end{titlepage}
\begin{center}
  {\Large        The OPAL Collaboration}
\end{center}\bigskip
\begin{center}{
G.\thinspace Abbiendi$^{  2}$,
C.\thinspace Ainsley$^{  5}$,
P.F.\thinspace {\AA}kesson$^{  3}$,
G.\thinspace Alexander$^{ 22}$,
J.\thinspace Allison$^{ 16}$,
P.\thinspace Amaral$^{  9}$, 
G.\thinspace Anagnostou$^{  1}$,
K.J.\thinspace Anderson$^{  9}$,
S.\thinspace Arcelli$^{  2}$,
S.\thinspace Asai$^{ 23}$
% ...
}\end{center}
\bigskip\bigskip
$^{  1}$School of Physics and Astronomy, University of Birmingham,
Birmingham B15 2TT, UK
\newline
$^{  2}$Dipartimento di Fisica dell' Universit\`a di Bologna and INFN,
I-40126 Bologna, Italy
\newline
$^{  3}$Physikalisches Institut, Universit\"at Bonn,
D-53115 Bonn, Germany
\newline
% ...
$^{  5}$Cavendish Laboratory, Cambridge CB3 0HE, UK
\newline
% ...
$^{  9}$Enrico Fermi Institute and Department of Physics,
University of Chicago, Chicago IL 60637, USA
\newline
% ...
$^{ 16}$Department of Physics, Schuster Laboratory, The University,
Manchester M13 9PL, UK
\newline
% ...
$^{ 22}$Department of Physics and Astronomy, Tel Aviv University,
Tel Aviv 69978, Israel
\newline
$^{ 23}$International Centre for Elementary Particle Physics and
Department of Physics, University of Tokyo, Tokyo 113-0033, and
Kobe University, Kobe 657-8501, Japan
\newline
% ...
$^{  a}$ and at TRIUMF, Vancouver, Canada V6T 2A3
\newline
$^{  b}$ and Royal Society University Research Fellow
\newline
% ...
$^{  d}$ and Heisenberg Fellow
\newline
$^{  e}$ and Department of Experimental Physics, Lajos Kossuth University,
 Debrecen, Hungary
\newline
% ...
$^{  h}$ now at University of Liverpool, Dept of Physics,
Liverpool L69 3BX, UK
\newline
$^{  i}$ and CERN, EP Div, 1211 Geneva 23
\newline
% ...
$^{  k}$ now at University of Kansas, Dept of Physics and Astronomy,
Lawrence, KS 66045, USA
\newline
$^{  l}$ now at University of Toronto, Dept of Physics, Toronto, Canada 
\newline
$^{  m}$ current address Bergische Universit\"at,  Wuppertal, Germany

%
%
\section{Introduction}
In this paper searches for new neutral scalar bosons \Szero with the
\OPAL detector at \LEP are described. The new bosons are assumed to be
produced in association with a \Zzero boson via the Bjorken process
$\ee \to \mathrm{\Szero{}\Zzero}$.  Throughout this note, \Szero
denotes, depending on the context, any new scalar neutral boson, the
Standard Model Higgs boson $\mathrm{H}^0_{\mathrm{SM}}$ or CP-even
Higgs bosons \hzero in models that predict more than one Higgs boson.

The analyses are topological searches and are based on studies of the
recoil mass spectrum in $\Zzero\to\ee$ and $\mu^+ \mu^-$ events and on
a search for $\Szero\Zzero$ events with $\Szero \to \ee$ or photons
and $\Zzero \to \nu\bar{\nu}$.  They are sensitive to all decays of
\Szero into an arbitrary combination of hadrons, leptons, photons and
invisible particles, and to the case of a long-lived \Szero leaving
the detector without interacting. The analyses are applied to 
\LEP~1 \Zzero on-peak data (115.4~pb$^{-1}$ at $\sqrt{s}=91.2~\GeV$)
and to 662.4~pb$^{-1}$ of \LEP~2 data collected at centre-of-mass
energies in the range of 183 to 209~\GeV. In 1990 \OPAL performed a
decay-mode independent search for light Higgs bosons and new scalars
using 6.8~pb$^{-1}$ of data with centre-of-mass energies around the
\Zzero pole \cite{c:Nguyen}.  Assuming the Standard Model production
cross section, a lower limit on the Higgs boson mass of $11.3~\GeV$
was obtained. We have re-analysed the \LEP~1 on-peak data in
order to extend the sensitive region to signal masses up to 55~\GeV.
Including the data above the \Zzero peak (\LEP~2) enlarges the
sensitivity up to $m_{\Szero}\sim 100~\GeV$. The \Szero mass range
between 30 and 55~\GeV is covered by both the \LEP~1 and the \LEP~2
analysis.

The results are presented in terms of limits on the scaling factor
\sq, which relates the \Szero{}\Zzero production cross section to the
Standard Model (\klein{SM}) cross section for the Higgs-strahlung
process:
\begin{equation}\label{e:sq_def}
  \sigma_{\mathrm{\Szero\Zzero}} = \sq\cdot
  \sigma_{\mathrm{\HSM\Zzero}}(m_{\mathrm{\HSM}}=m_{\mathrm{\Szero}}),
\end{equation} 
where we assume that \sq does not depend on the centre-of-mass energy
for any given mass $m_{\mathrm{\Szero}}$. Since the analysis is
insensitive to the decay mode of the \Szero, these limits can be
interpreted in any scenario beyond the Standard Model. Examples of
such interpretations are listed in the following.

\begin{itemize} 
\item The most general case is to provide upper limits on the
  cross section or scaling factor \sq for a single new scalar boson
  independent of its couplings to other particles. We assume that the
  decay width is small compared to the detector mass resolution.  In a
  more specific interpretation, assuming the \Szero{}\Zzero production
  cross section to be identical to the Standard Model Higgs boson
  one, the limit on \sq can be translated into a lower limit on the
  Higgs boson mass\footnote{Dedicated searches for the Standard Model
    Higgs boson by the four \LEP experiments, exploiting the
    prediction for its decay modes, have ruled out masses of up to
    114.1~\GeV \cite{c:LEP_Higgs_limit}.}.
    
\item For the first time we give limits not only for a single mass
  peak with small width, but also for a continuous distribution of the
  signal in a wide mass range. Such continua appear in several
  recently proposed models, \emph{e.\,g.} for a large number of
  unresolved Higgs bosons about equally spaced in mass (``Uniform Higgs
  scenario''~\cite{c:gunion}), or models with additional
  SU(3)$_{\mathrm{C}}
  \times$SU(2)$_{\mathrm{L}}\times$U(1)$_{\mathrm{Y}}$ singlet fields
  which interact strongly with the Higgs boson (``Stealthy Higgs
  scenario''~\cite{c:stealthy_higgs}). These two models are described in
  more detail in the next section.
\end{itemize}

\section{Continuous Higgs scenarios}
\subsection{The Uniform Higgs scenario}\label{worst-case}
This model, as described in Ref.~\cite{c:gunion}, assumes a broad
enhancement over the background expectation in the $M_{\mathrm{X}}$
mass distribution for the process $\ee\to \Zzero\mathrm{X}$.  This
enhancement is due to numerous additional neutral Higgs bosons \hi
with masses $m_{\mathrm{A}} \le m(\hi) \le m_{\mathrm{B}}$, where
$m_{\mathrm{A}}$ and $m_{\mathrm{B}}$ indicate the lower and upper
bound of the mass spectrum.  The squared coupling, $g^2$, of the Higgs states \hi
to the \Zzero is modified by a factor $k_i$ compared to the \SM
\Hzero{}\Zzero coupling: $g^2_{\Zzero\hi} = k_i\cdot
g^2_{\Zzero\hzero_{\mathrm{SM}}}$.
  
If the Higgs states are assumed to be closer in mass than the
experimental mass resolution, then there is no need to distinguish
between separate $k_i$.  In this case the Higgs states and their
reduction factors $k_i$ can be combined into a coupling density
function, $\Ktilde(m) = \mathrm{d}k/\mathrm{d}m$.  The model obeys two
sum rules which in the limit of unresolved mass peaks can be expressed
as integrals over this coupling density function:
\begin{eqnarray} 
  \int\limits_{0}^{\infty} \mathd\mhwo\; \tilde{K}(\mhwo) & = & 1 \label{eq:sumrule1}\\ 
  \int\limits_{0}^{\infty} \mathd\mhwo\; \tilde{K}(\mhwo){\mhwo}^2 & \le & m^2_\mathrm{C}, 
  \label{eq:sumrule2} 
\end{eqnarray}
where $\Ktilde(m) \ge 0$ and \mC is a perturbative mass scale of the
order of 200~\GeV.  The value of \mC is model dependent and can be
derived by requiring that there is no Landau pole up to a scale
$\Lambda$ where new physics occurs \cite{c:gunion}.  If neither a
continuous nor a local excess is found in the data,
Equation~\ref{eq:sumrule1} can be used to place constraints on the
coupling density function $\Ktilde(m)$. For example, if $\Ktilde(m)$
is assumed to be constant over the interval [\mA, \mB{}] and zero
elsewhere,
\begin{eqnarray}
  \Ktilde(m) & = & 1/\left(\mB-\mA\right) \quad \mathrm{for}~\mA \le m \le \mB,\nonumber\\
             & = & 0 \quad \mathrm{elsewhere,}\nonumber
\end{eqnarray}
then certain choices for the interval [\mA, \mB{}] can be excluded. From
this and from Equation~\ref{eq:sumrule2} lower limits on the mass
scale \mC can be derived.

\subsection{The Stealthy Higgs scenario}
This scenario predicts the existence of additional SU(3)$_{\mathrm{C}}
\times$SU(2)$_{\mathrm{L}}\times$U(1)$_{\mathrm{Y}}$ singlet fields
(phions), which would not interact via the strong or electro-weak
forces, thus coupling only to the Higgs boson \cite{c:stealthy_higgs}.
Therefore these singlets would reveal their existence only in the
Higgs sector by offering invisible decay modes to the Higgs boson. The
width of the Higgs resonance can become large if the number of such
singlets, $N$, or the coupling $\omega$ is large, thus yielding a
broad spectrum in the mass recoiling against the reconstructed \Zzero.
The interaction term between the Higgs and the additional phions in
the Lagrangian is given by
\begin{equation}
  \mathscr{L}_{\mathrm{interaction}} = 
    -\frac{\omega}{2\sqrt{N}}
    \vec{\varphi}^2\phi^\dagger\phi,
\end{equation}
where $\phi$ is the Standard Model Higgs doublet, $\omega$ is the
coupling constant, and $\vec{\varphi}$ is the vector of the new
phions. An analytic expression for the Higgs width can be found in the
limit $N\to\infty$:
\begin{equation}\label{eq:higgs_width}
  \Gamma_{\mathrm{H}}(\mHsm) = \Gamma_{\mathrm{SM}}(\mHsm) + \frac{\omega^2 v^2}
                  {32\, \pi\, \mHsm},
\end{equation}
where $v$ is the vacuum expectation value of the Higgs field.  This
expression results when setting other model parameters to zero,
including the mass of the phions \cite{c:stealthy_higgs}.  The
cross section for the Higgs-strahlung process can be calculated from
Equations~9 and 10 of reference~\cite{c:stealthy_higgs}.

In section~\ref{s:stealthy} we derive limits on the Stealthy Higgs
model which can be compared to expected limits from dedicated
$\Hzero\to$ invisible searches, which are estimated in
Ref.~\cite{c:stealthy_higgs} for the same scenario.  By simulating
signal spectra for different Higgs widths $\Gamma_{\mathrm{H}}$ we set
limits in the $\omega$-$m_\mathrm{H}$ plane in the large $N$ limit.

%
%
\section{Data sets and Monte Carlo samples}
The analyses are based on data collected with the \OPAL detector at
\LEP during the runs in the years 1991 to 1995 at the \Zzero peak
(\LEP~1) and on data taken in the years 1997 to 2000 at centre-of-mass
energies between 183 and 209~\GeV (\LEP~2).  The integrated luminosity
used is 115.4~pb$^{-1}$ for the \LEP~1 energy and 662.4~pb$^{-1}$ for
the \LEP~2 energies, as detailed in Table~\ref{t:lumi}. A description
of the \OPAL detector\footnote{OPAL uses a right handed coordinate
  system. The $z$ axis points along the direction of the electron beam
  and the $x$ axis is horizontal pointing towards the centre of the
  LEP ring. The polar angle $\theta$ is measured with respect to the
  $z$ axis, the azimuthal angle $\phi$ with respect to the $x$ axis.}
can be found elsewhere \cite{c:detector}.

To estimate the detection efficiency for a signal from a new scalar
boson and the amount of background from \sm processes, several \mc
samples are used.  Signal events are simulated for masses from 1~\keV
to 110~\GeV in a large variety of decay modes with the
\klein{HZHA}~\cite{c:MC_HZHA} generator. The signal efficiencies are
determined for all possible decays of a \SM Higgs boson (quarks,
gluons, leptons, photons), for the decays into `invisible' particles
(e.\,g.  Lightest Supersymmetric Particles) $\Szero\to\chi^0\chi^0$ as
well as for `nearly invisible' decays, $\Szero\to\chi^0_2\chi^0_1$,
where the $\chi^0_2$ decays into a $\chi^0_1$ plus a photon or a
virtual \Zzero, and for decays $\Szero \to \mathrm{AA}$ with A$\to$
cc, gg or $\tau\tau$, where A is the CP-odd Higgs boson in
supersymmetric extensions of the \SM. For simulation of background
processes the following generators are used:
\klein{BHWIDE}~\cite{c:MC_BHWIDE}, \klein{TEEGG}~\cite{c:MC_TEEGG}
(($\mathrm{Z}/\gamma)^*\to\mathrm{e}^+\mathrm{e}^-(\gamma)$),
\klein{KORALZ}~\cite{c:MC_KORALZ}, \klein{KK2F}~\cite{c:MC_KK2F} (both
$\mu^+\mu^-(\gamma)$ and $\tau^+\tau^-(\gamma)$),
\klein{JETSET}~\cite{c:MC_PYTHIA}, \klein{PYTHIA}~\cite{c:MC_PYTHIA}
(q$\bar{\mathrm{q}}(\gamma)$), \klein{GRC4F}~\cite{c:MC_GRC4F}
(four-fermion processes), \klein{PHOJET}~\cite{c:MC_PHOJET},
\klein{HERWIG}~\cite{c:MC_HERWIG}, Vermaseren~\cite{c:MC_VERMASEREN}
(hadronic and leptonic two-photon processes),
\klein{NUNUGPV}~\cite{c:MC_NUNUGPV} ($\nu\bar{\nu}\gamma$) and
\klein{RADCOR}~\cite{c:MC_RADCOR} ($\gamma\gamma$).  For all Monte
Carlo generators other than \klein{HERWIG}, the hadronisation is done
using \klein{JETSET}.  The luminosity of the main background Monte
Carlo samples is at least 4 times the statistics of the data for the
two-fermion background, 50 times for the four-fermion background and 5
times for the two-photon background. The signal Monte Carlo samples
contain 500--1000 events per mass and decay mode.  The generated
events are passed through a detailed simulation of the \OPAL
detector~\cite{c:GOPAL} and are reconstructed using the same
algorithms as for the real data.

\section{Decay-mode independent searches for 
\boldmath e$^+$e$^-\to$S$^0$Z$^0$\unboldmath}
\label{s:decay_independent_searches}
The event selection is intended to be efficient for the complete
spectrum of possible \Szero decay modes.  As a consequence it is
necessary to consider a large variety of background processes.
Suppression of the background is performed using the smallest amount
of information possible for a particular decay of the \Szero.  The
decays of the \Zzero into electrons and muons are the channels with
highest purity, and therefore these are used in this analysis. They
are referred to as the electron and the muon channel, respectively.
The signal process can be tagged by identifying events with an
acoplanar, high momentum electron or muon pair.  
%
We use the term `acoplanar' for lepton pairs if the two leptons and
the beam axis are not consistent with lying in a single plane.

Different kinematics of the processes in the \LEP~1 and the \LEP~2
analysis lead to different strategies for rejecting the background.
At \LEP~2 the invariant mass of the two final-state leptons in the
signal channels is usually consistent with the \Zzero mass, while this
is not true for a large part of the background. Therefore a cut on the
invariant mass rejects a large amount of background. Remaining
two-fermion background from radiative processes can partially be
removed by using a photon veto without losing efficiency for photonic
decays of the \Szero.  In the \LEP~1 analysis the invariant mass of
the lepton pair cannot be constrained. Therefore, stronger selection
cuts have to be applied to suppress the background, resulting in an
insensitivity to the decays $\Szero \to $ photons and at low masses
also to $\Szero \to \ee$.  Hence, these decay modes are recovered in a
search dedicated to $\ee\to \Szero\Zzero$ with $\Zzero \to
\nu\bar{\nu}$ and $\Szero \to$ photons (or photons plus invisible
particles) or electrons at low masses $\mS < 500~\MeV$.

\subsection{Event selection for \boldmath$\ee\to$ S$^0$Z$^0$ with
  $\Zzero\to \ee$ or $\mm$}\label{s:LEP1_cl} 

The analysis starts with a preselection of events that contain at
least two charged particles identified as electrons or muons.  A
particle is identified as an electron or muon, if it is identified by
at least one of the two methods:
\begin{itemize}
\item The standard \OPAL procedures for electron and muon
  identification~\cite{c:IDpackage}. These routines were developed to
  identify leptons in a hadronic environment. Since the signal events
  contain primarily isolated leptons, a second method with a higher
  efficiency is also used:
\end{itemize}
\begin{itemize}
\item A track is classified as an electron if the ratio $E/p$ is
  greater than 0.8, where $p$ is the track momentum and $E$ the
  associated electromagnetic energy.  Furthermore the energy loss
  d$E$/d$x$ in the central tracking chamber has to be within the
  central range of values where 99\,\% of the electrons with this
  momentum are expected.  Muons are required to have $E/p < 0.2$ and
  at least three hits in total in the muon chambers plus the last
  three layers of the hadronic calorimeter.
\end{itemize}
The two tracks must have opposite charge and high momentum.  Depending
on the recoil mass of the lepton pair, the \LEP~1 analysis requires a
momentum of the higher energy lepton above 20--27~\GeV in the electron
channel and above 20--30~\GeV in the muon channel. The momentum of the
lower energy lepton has to be greater than 10--20~\GeV in both
channels.

For electrons these cuts apply to the energy deposited in the
electromagnetic calorimeter, and for the muons to the momentum
measured in the tracking system.  At \LEP~2 energies the lepton
momenta have a weaker dependence on the recoil mass, therefore fixed
cuts are used which are adjusted for the different centre-of-mass
energies: $E_1 > 0.22\cdot\sqrts$, $E_2 > 0.11\cdot\sqrts$\quad for
electrons and $p_1 > 0.22\cdot\sqrts$, $p_2 > 0.12\cdot\sqrts$\quad
for muons, where $E_1$, $p_1$ and $E_2$, $p_2$ are the energy and
momentum of the lepton with the higher and lower momentum,
respectively.

The two leptons must be isolated from the rest of the event.  The
isolation angle \aiso of a lepton candidate is defined as the maximum
angle for which the energy $E_{\mathrm{cone}}$ contained within a cone
of half-angle \aiso around the direction of the lepton at the vertex
is less than 1~\GeV.  $E_{\mathrm{cone}}$ is the energy of all tracks
and electromagnetic clusters not associated to a track within the
cone, excluding the energy of the lepton itself.  Leptons at small
angles to the beam axis ($|\cos\theta| > 0.9$ in the electron channel
and $|\cos\theta| > 0.94$ in the muon channel) are not used due to
detector inefficiencies and mismodelling in this region.  These cuts
also serve to reduce the background from two-fermion and two-photon
processes.  We ignore lepton candidates inside a \degree{0.3}
azimuthal angle to the anode planes of the jet chamber since they are
not well described in the detector simulation.  If more than one
electron or muon pair candidate with opposite charge is found, for the
\LEP~1 analysis the two leptons with the highest momentum, and for the
\LEP~2 analysis the pair with invariant mass closest to $m_{\Zzero}$
are taken as \Zzero decay products.

The background to the \Szero$\ell^+\ell^-$ signal arises from several
processes which are suppressed as described below:

\begin {itemize}
\item In $(\mathrm{Z}/\gamma)^*\to\ee,\mm$ events without initial or
  final state radiation the leptons are produced in a back-to-back
  topology. We reject these events by cutting on the acoplanarity
  angle \pa which is defined as $\pi - \phi_{\mathrm{open}}$, where
  $\phi_{\mathrm{open}}$ is the opening angle between the two lepton
  tracks in the plane perpendicular to the beam axis. For the \LEP~1
  analysis the acoplanarity angle is multiplied by the average of the
  $\sin\theta$ of the tracks in order to account for the larger
  influence of the track direction resolution on the acoplanarity
  angle for tracks with small $\sin\theta$.  The modified acoplanarity
  angle is termed $\alpha$. The cuts are $0.11~\mathrm{rad} < \alpha
  < 2.0$~rad and \mbox{$\pa > 0.15$---$0.20$~rad} (depending on the
  centre-of-mass energy).
        
\item In two-photon processes, where the incoming electron and positron
  are scattered at low angles, usually one or both of the
  electrons are undetected. Events of this type usually have large
  missing momentum with the missing momentum vector,
  $\vec{p}_{\mathrm{miss}}$, pointing at low angles to the beam axis.
  In $(\mathrm{Z}/\gamma)^*\to\ee,\mm$ events with initial-state
  radiation the photons usually remain undetected at low angles.  The
  requirement $|\cos\tpmiss| < 0.98$ for the \LEP~1 analysis and
  $|\cos\tpmiss| < 0.95$ for the \LEP~2 analysis reduces background
  from these two sources.
  
\item The semileptonic decays of b- or c-mesons provide another source
  of leptons which can be misidentified as direct \Zzero decay
  products. This background is reduced by requiring the leptons to be
  isolated from the rest of the event.  We require one of the
  isolation angles of the two lepton candidates to be greater than
  \degree{20} and the other one to be greater than \degree{10} for the
  \LEP~1 analysis and to be greater than \degree{15} and \degree{10},
  respectively, for the \LEP~2 analysis.
\end {itemize}

\noindent Up to this point the analyses for \LEP~1 and \LEP~2
energies are essentially identical, but they are tuned separately, as
detailed in Table~\ref{t:cuts_lep1}.  The different features of signal
and background at \LEP~1 and \LEP~2 energies are taken into account
with the following cuts.

\subsubsection{Cuts used only in the \LEP~1 selection}
\label{s:lep1-selection}
\begin {itemize}
\item Since the electron or muon pair originates from a \Zzero its
  invariant mass is high in comparison to a typical pair of isolated
  leptons in hadronic background. We therefore require the lepton pair
  invariant mass to exceed 20~\GeV.
  
\item At this stage the cut selection is still sensitive to all decay
  modes of the \Szero. The main background, however, arises from
  electron and muon pairs accompanied by energetic photon radiation.
  Reduction of this kind of background is made by applying cuts on
  photons and electrons recognised as coming from
  a photon conversion.\\
  \hspace*{2ex} Events with less than four tracks are vetoed if there
  is an unassociated cluster in the electromagnetic calorimeter with
  an energy greater than 1~\GeV outside a \degree{10} cone around a
  lepton candidate (photon veto).  They are also vetoed if the energy
  in the forward calorimeters, corresponding to the polar angle region
  47--200 mrad, exceeds 2~\GeV (forward veto).  In order to reject
  events where the photon converts into an electron-positron pair,
  events with one, two or three tracks in addition to the lepton are
  excluded if at least one of them is identified as a track from a
  conversion (conversion veto). The conversion finder is based on an
  artificial neural network \cite{c:conversion}.\\
  \hspace*{2ex} The photon and the conversion veto are at the expense
  of sensitivity for decays $\Szero \to$ photons (or photons plus
  invisible particles) in the whole mass region and for $\Szero\to\ee$
  at low masses ($m_{\Szero} < 500~\MeV$). In order to retain
  sensitivity to these decay modes, they are taken into account in a
  search dedicated to $\ee\to \Szero\Zzero$ with $\Zzero \to
  \nu\bar{\nu}$ and $\Szero \to$ photons (or photons plus invisible
  particles) and for $\mS < 500~\MeV$ also to electrons as described
  in section~\ref{s:neutralchannel}.
\end {itemize}

\begin{figure}
\centering
\includegraphics[width=16cm]{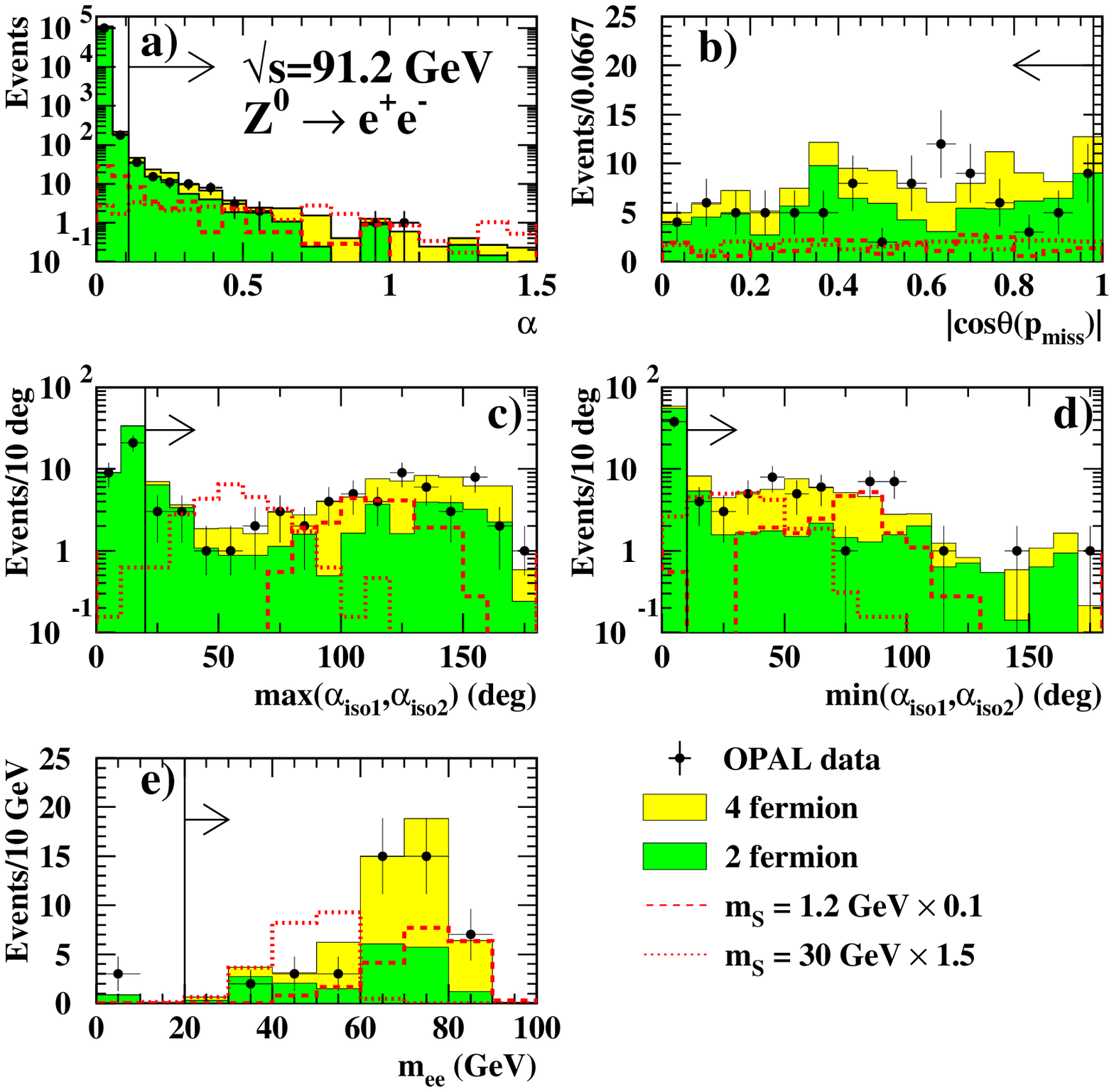}

\caption{\label{f:cutvars_eeLEP1}
  Cut variables for $\Zzero \to \ee$ at $\sqrt{s}$ = 91.2~GeV. The
  \klein{OPAL} data are indicated by dots with error bars (statistical
  error), the four-fermion background by the light grey histograms and
  the two-fermion background by the medium grey histograms. The signal
  distributions from a 1.2~\GeV \Szero are plotted as dashed lines and
  those from a 30~\GeV \Szero as dotted lines, respectively. The
  signal histograms are normalised corresponding to 0.1 and 1.5 times
  of the \SM Higgs-strahlung cross section and show the decay channel
  $\Szero \to \mathrm{gg}$.  Each variable is shown with the cuts
  applied before the cut on this variable is done, respecting the
  order of cuts in Table~\ref{t:cutflow_LEP1}.  The arrows indicate
  the accepted regions. The histograms in Figure a) have non-constant
  bin widths.
  }
\end{figure}

\begin{figure}

\centering
\includegraphics[width=16cm]{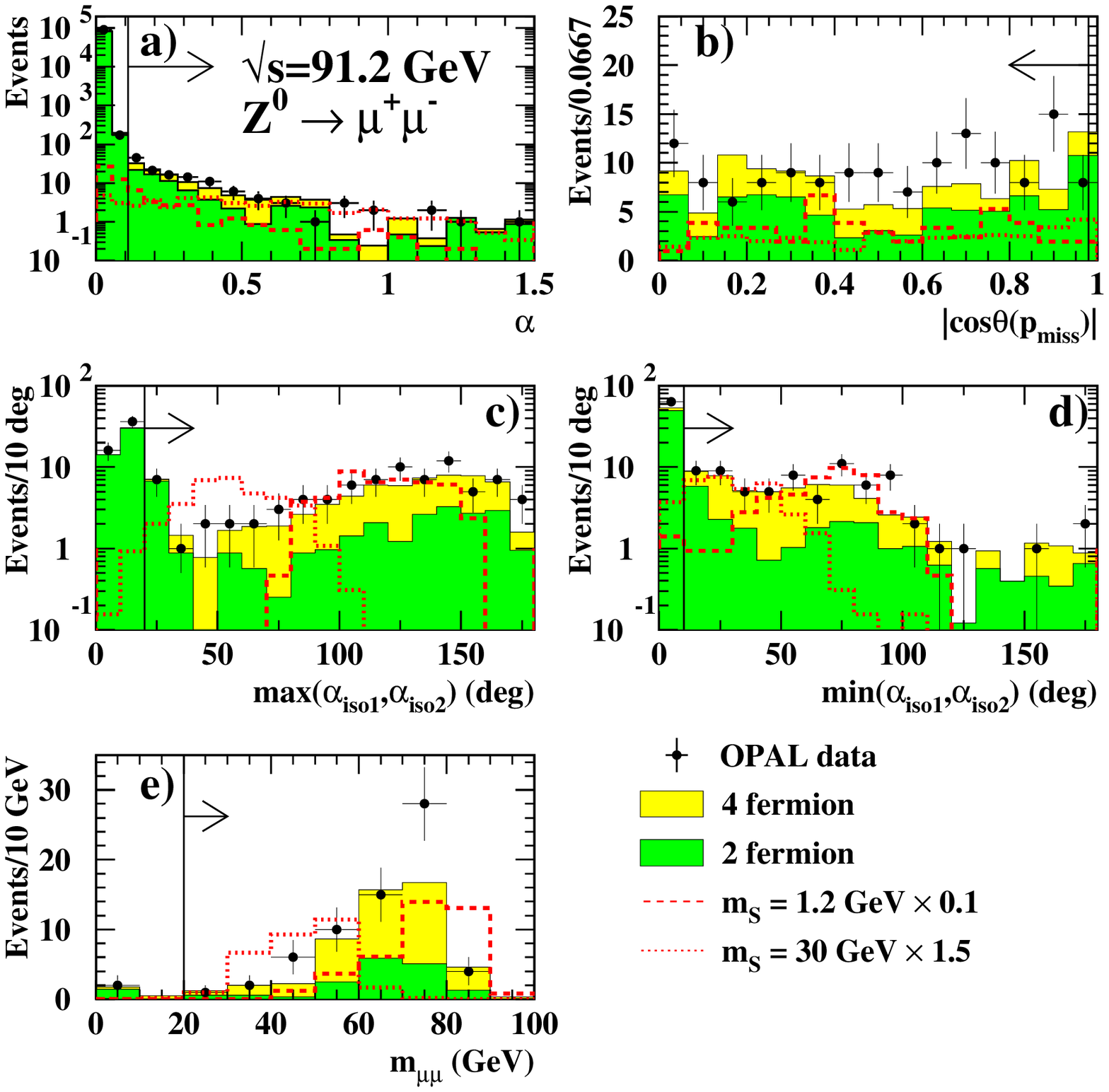}

\caption{\label{f:cutvars_mmLEP1}
  Cut variables for $\Zzero \to \mm$ at $\sqrt{s}$ = 91.2~GeV. The
  \klein{OPAL} data are indicated by dots with error bars (statistical
  error), the four-fermion background by the light grey histograms and
  the two-fermion background by the medium grey histograms. The signal
  distributions from a 1.2~\GeV \Szero are plotted as dashed lines and
  those from a 30~\GeV \Szero as dotted lines, respectively. The
  signal histograms are normalised corresponding to 0.1 and 1.5 times
  of the \SM Higgs-strahlung cross section and show the decay channel
  $\Szero \to \mathrm{gg}$.  Each variable is shown with the cuts
  applied before the cut on this variable is done, respecting the
  order of cuts in Table~\ref{t:cutflow_LEP1}.  The arrows indicate
  the accepted regions. The histograms in Figure a) have non-constant
  bin widths.
  }
\end{figure}

All cuts are listed in Table~\ref{t:cuts_lep1} and the number of
events after each cut is given in Table~\ref{t:cutflow_LEP1}. The
distributions of the cut variables in data and Monte Carlo are shown
in Figures~\ref{f:cutvars_eeLEP1} and \ref{f:cutvars_mmLEP1}. After
the selection 45 events remain in the channel $\Zzero \to \ee$, with
\err{55.2}{3.0}{3.0} events expected from \klein{SM} background (the
evaluation of the systematic uncertainties is described in
section~\ref{s:ee,mm_systematics}). In the channel $\Zzero \to
\mu^+\mu^-$, 66 events remain in the data with \err{53.6}{2.7}{2.1}
expected from \klein{SM} background.

\begin{figure}
  \centering
  \includegraphics[width=0.48\textwidth]{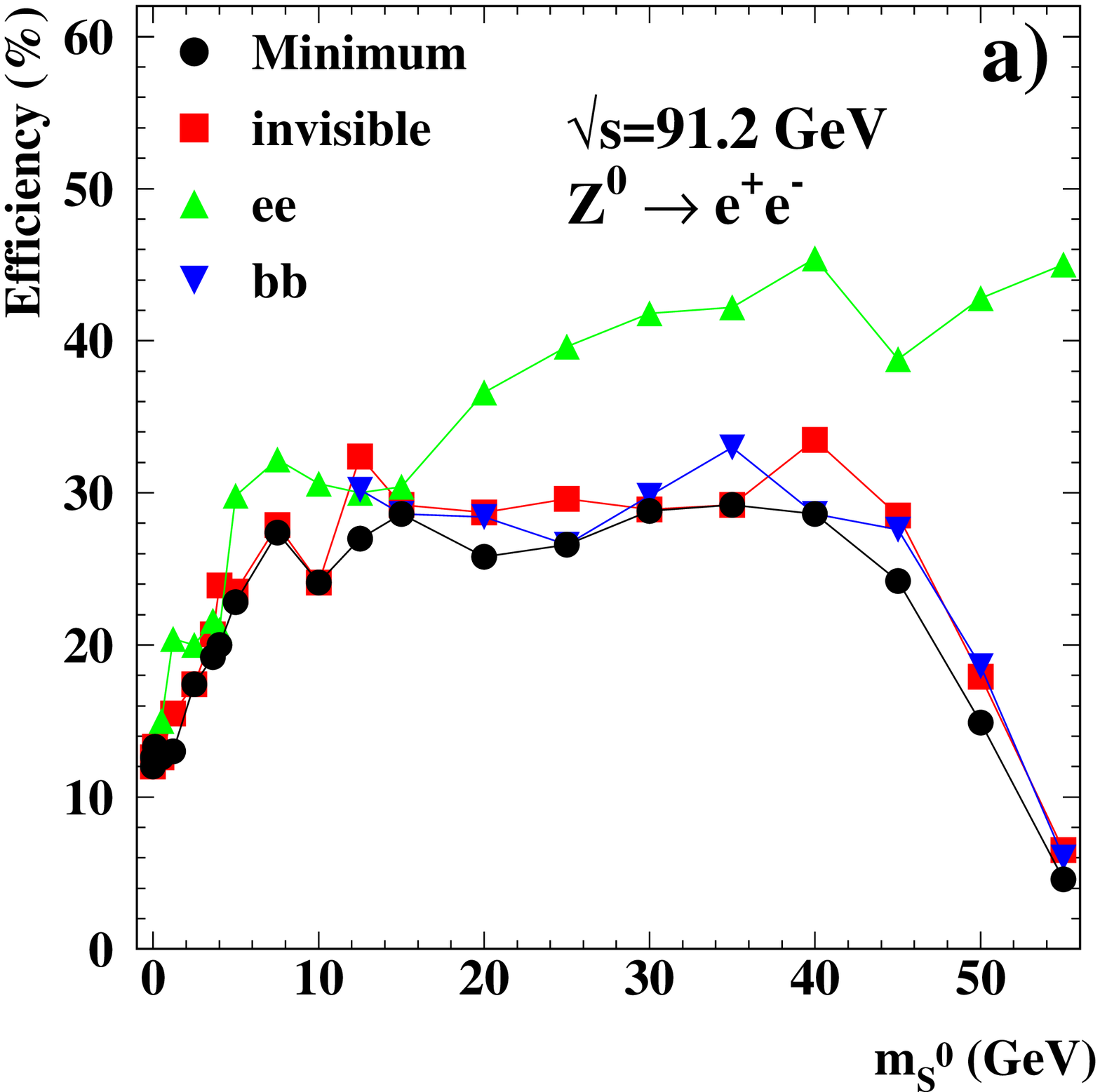}
  \includegraphics[width=0.48\textwidth]{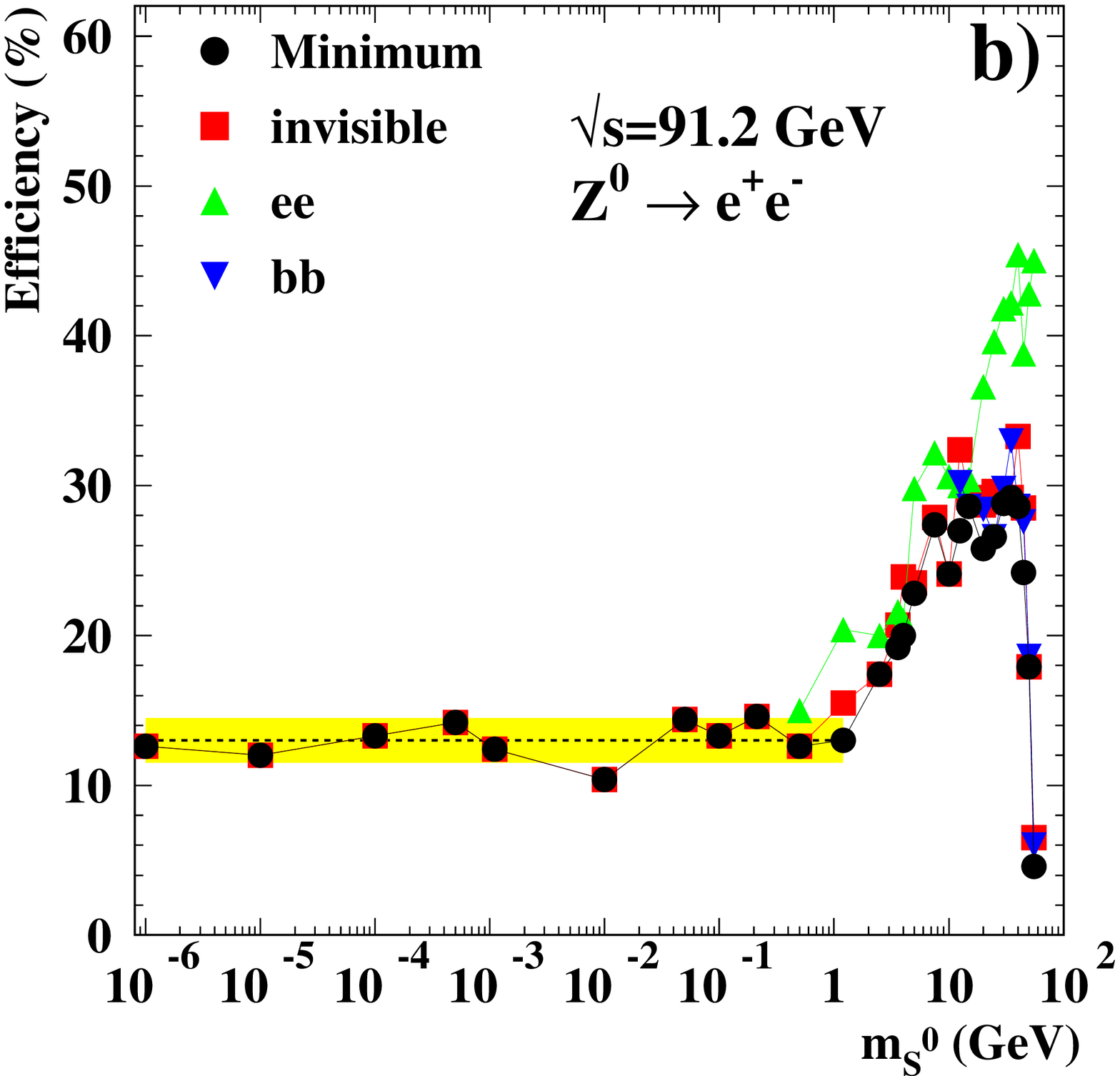}\\[4ex]
  \includegraphics[width=0.48\textwidth]{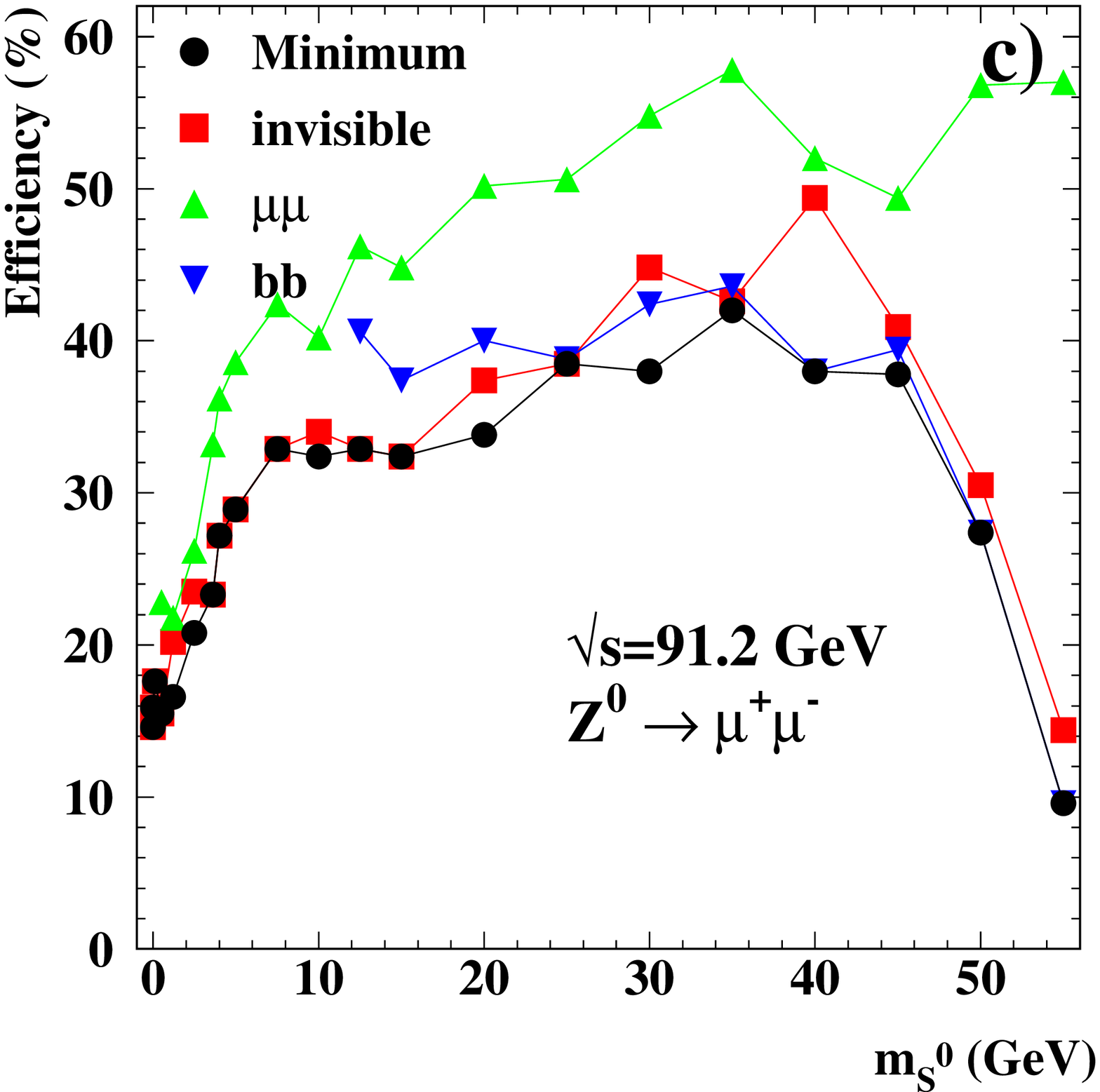}
  \includegraphics[width=0.48\textwidth]{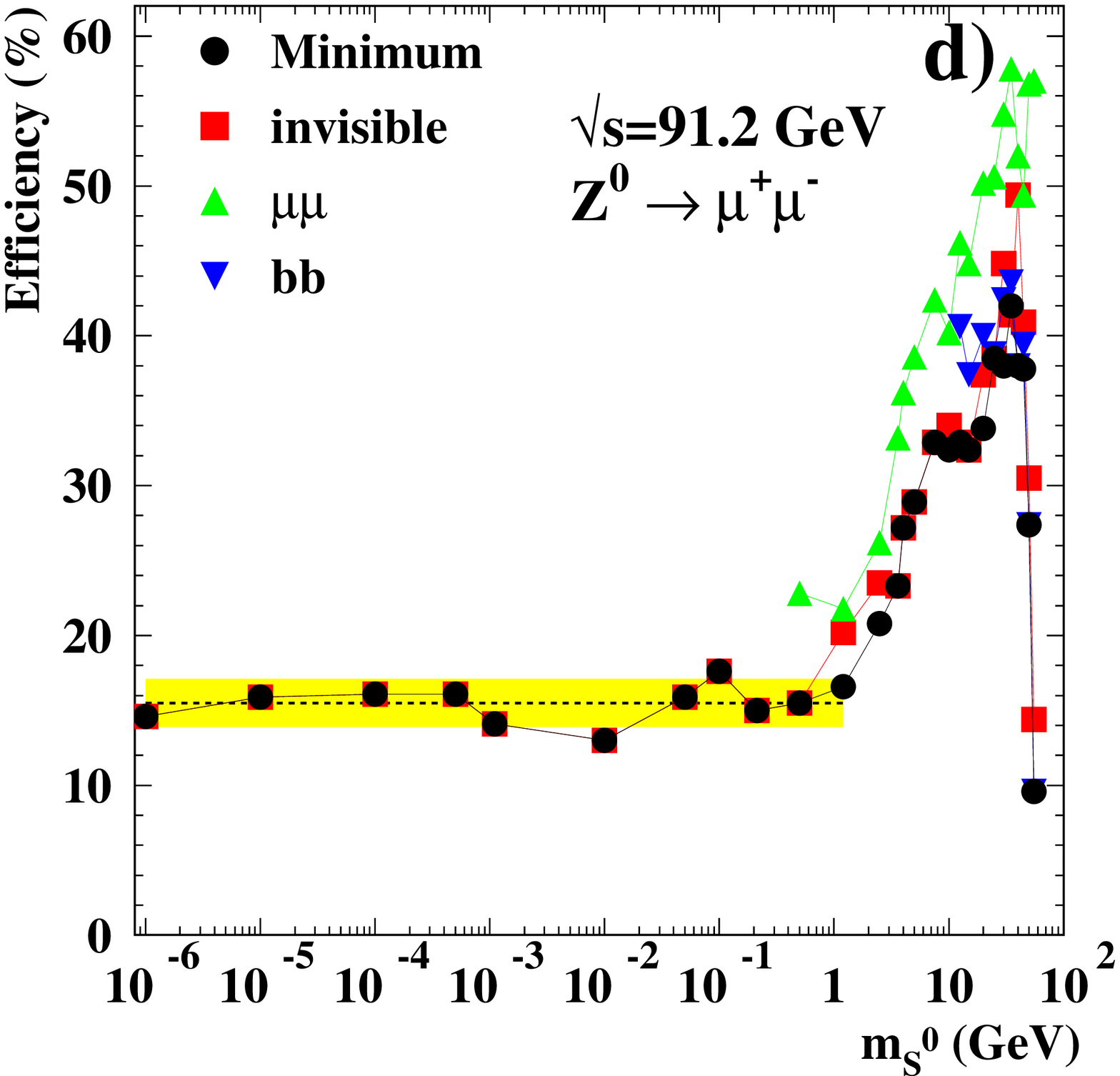}
  \caption{\label{f:eff91}
    The efficiency versus the \Szero mass at $\sqrt{s}=91.2$~GeV for a
    subset of decay modes of \Szero: a)+b)~$\Zzero\to\ee$ in linear
    and logarithmic mass scale; c)+d)~$\Zzero\to\mm$ in linear and
    logarithmic mass scale. The minimum efficiencies which are used in
    the limits are also shown. In the low mass region, below the
    threshold for the decays of the \Szero into a pair of SM fermions,
    only the decays into photons or invisible particles are possible.
    For $\mS \lesssim \Gamma_{\Zzero}$ the efficiency is almost flat.
    This is indicated by the dashed line which marks the average
    efficiency for $\mS \le 1$~\GeV. The shaded bands show the typical
    error on the efficiencies in this region.  
    }
\end{figure}

\begin{figure}
  \centering
  \includegraphics[width=15cm]{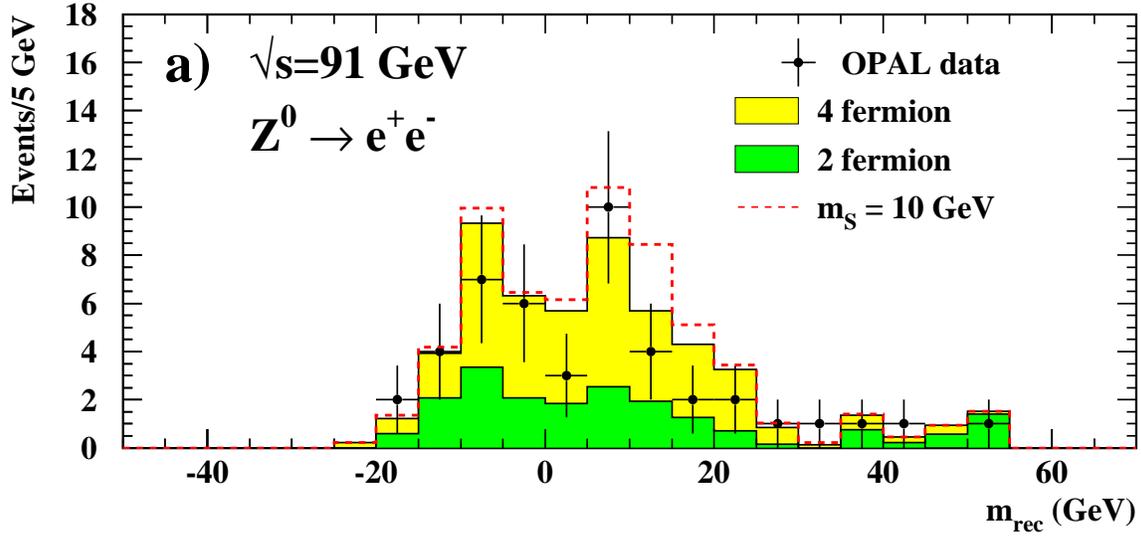}\\[5ex]
  \includegraphics[width=15cm]{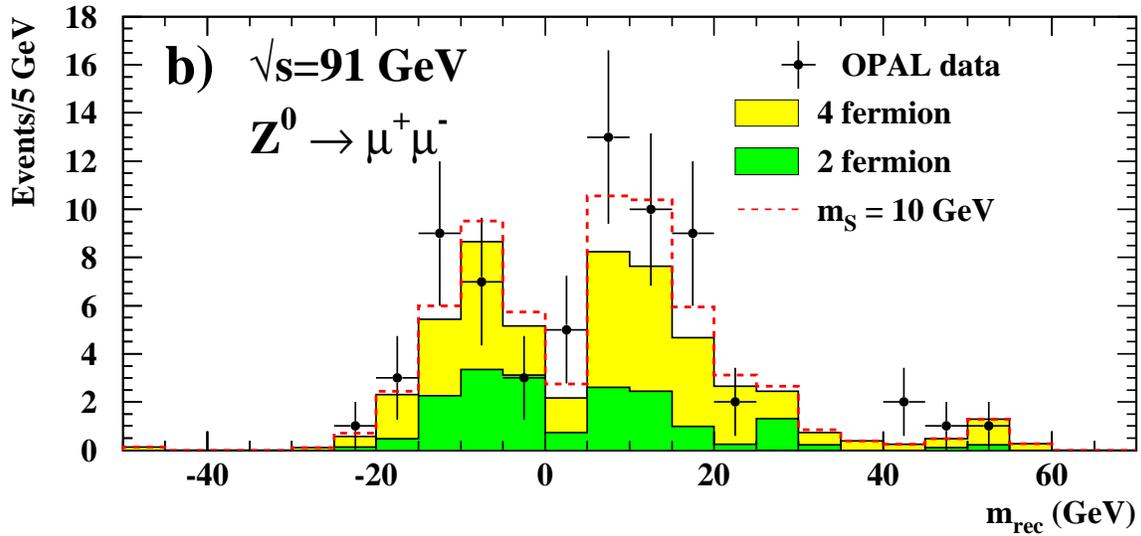}
\caption{\label{f:summass_LEP1} 
  The recoil mass spectra from $\sqrt{s}=$91.2~GeV a) for the decays
  $\Zzero \to \ee$ and b) for $\Zzero \to \mm$. \klein{OPAL} data are
  indicated by dots with error bars (statistical error), the
  four-fermion background by the light grey histograms and the
  two-fermion background by the dark grey histograms. The dashed lines
  for the signal distributions are plotted on top of the background
  distributions with normalisation corresponding to the cross section
  excluded at 95\% confidence level from the combination of both
  channels.
  }
\end{figure}

Figure~\ref{f:eff91} shows the efficiency versus the \Szero mass for
some example decay modes. The signal efficiency is at least 20\% in
the electron channel and at least 27\% in the muon channel for \Szero
masses between 4 and 45 GeV.  At masses below the kinematic threshold
for the decay of the \Szero into \ee ($\sim 1~\MeV$) only decays into
photons or invisible particles are possible. For each \Szero mass
hypothesis the smallest efficiency of all decay channels studied (also
shown in Figure~\ref{f:eff91}) is used in the limit calculation. The
analysis is sensitive to a large range of \Szero masses, down to
masses $m_{\Szero}$ well below $\Gamma_{\Zzero}$, where the
cross section increases significantly.  For this mass range mainly
soft bosons \Szero with energy $E_{\Szero} < \Gamma{_\Zzero}$ are
emitted, but the spectrum exhibits a significant tail to large
energies, which yields a detectable event topology.
Figure~\ref{f:summass_LEP1} shows the recoil mass spectrum to the
\Zzero decay products for both channels at \mbox{$\sqrt{s}$ =
  91.2~\GeV}.  The recoil mass squared is calculated from
\begin{equation}\label{eq:recmass}
m_{\mathrm{rec}}^2 
  = \left(\sqrt{s} - E_{\ell\ell}\right)^2 - p_{\ell\ell}^2,
\end{equation}
where $E_{\ell\ell}$ and $p_{\ell\ell}$ are the energy and the
momentum sum of the two lepton tracks, and $\sqrt{s}$ is the
centre-of-mass energy.  The momentum sum is calculated from the track
momentum of the \Zzero decay products in the muon channel and from the
track momentum and energy deposition of the electrons in the
electromagnetic calorimeter in the electron channel\footnote{Due to
  the limited energy and momentum resolution, the calculated value of
  $m_{\mathrm{rec}}^2$ can be negative.  We define $m_{\mathrm{rec}} =
  \sqrt{m_{\mathrm{rec}}^2}$ for $m_{\mathrm{rec}}^2 \ge 0$ and
  $m_{\mathrm{rec}} = -\sqrt{-m_{\mathrm{rec}}^2}$ for
  $m_{\mathrm{rec}}^2 < 0$.}.

\subsubsection{Cuts used only in the \LEP~2 selection}
In the analysis for \LEP~2 energies, signal and background
characteristics differ significantly from those at \LEP~1.\medskip

\begin {itemize} 
\item The most important difference compared to the \LEP~1 analysis is
  the fact that in signal processes an on-shell \Zzero boson is
  produced. The selection requires the invariant mass $m_{\ell\ell}$
  of the lepton pair to be consistent with the \Zzero mass.  Due to
  the limited detector mass resolution, invariant masses within
  $m_{\Zzero} \pm 8~\GeV$ and $m_{\Zzero} \pm 10~\GeV$ are accepted
  for the electron and the muon channel, respectively.
  
\item The dominant background at this stage originates from leptonic
  \Zzero decays with photon radiation in the initial state.  At
  centre-of-mass energies above $m_{\Zzero}$ the cross section for
  radiating one (or more) high energy initial-state photon(s) is
  enhanced if the effective centre-of-mass energy of the
  electron-positron pair after photon emission is close to the \Zzero
  mass.  Such events are called `radiative returns' to the \Zzero
  pole. These background events are characterised by an acolinear and
  sometimes acoplanar lepton pair and one or more high energy photons.
  Such events are rejected by a $\gamma$-veto: if there is only one
  cluster in the electromagnetic calorimeter not associated to a track
  and the energy $E_{\mathrm{unass}}$ of the cluster exceeds 60~\GeV,
  then the event is rejected.  Events with two tracks and more than
  3~\GeV energy deposition in the forward calorimeters (covering the
  polar angle region 47--200~mrad) are also vetoed.
  The cross section for two fermion production is much smaller at
  \LEP~2 than at \LEP~1 so events with final state radiation are not
  such an important background as in the \LEP~1 case.
  
\item In the remaining background from two-photon processes and
  $(\mathrm{Z}/\gamma)^*\to\ee,\mm$ with initial-state radiation the
  leptons carry considerable momentum along the beam axis.  We reject
  these events by requiring $|p^z_1+p^z_2| < 50~\GeV$ where $p^z_i$
  are the $z$-components of the momentum of the two lepton candidates.
\end{itemize}
  
\begin{figure}
\centering
\includegraphics[width=16cm]{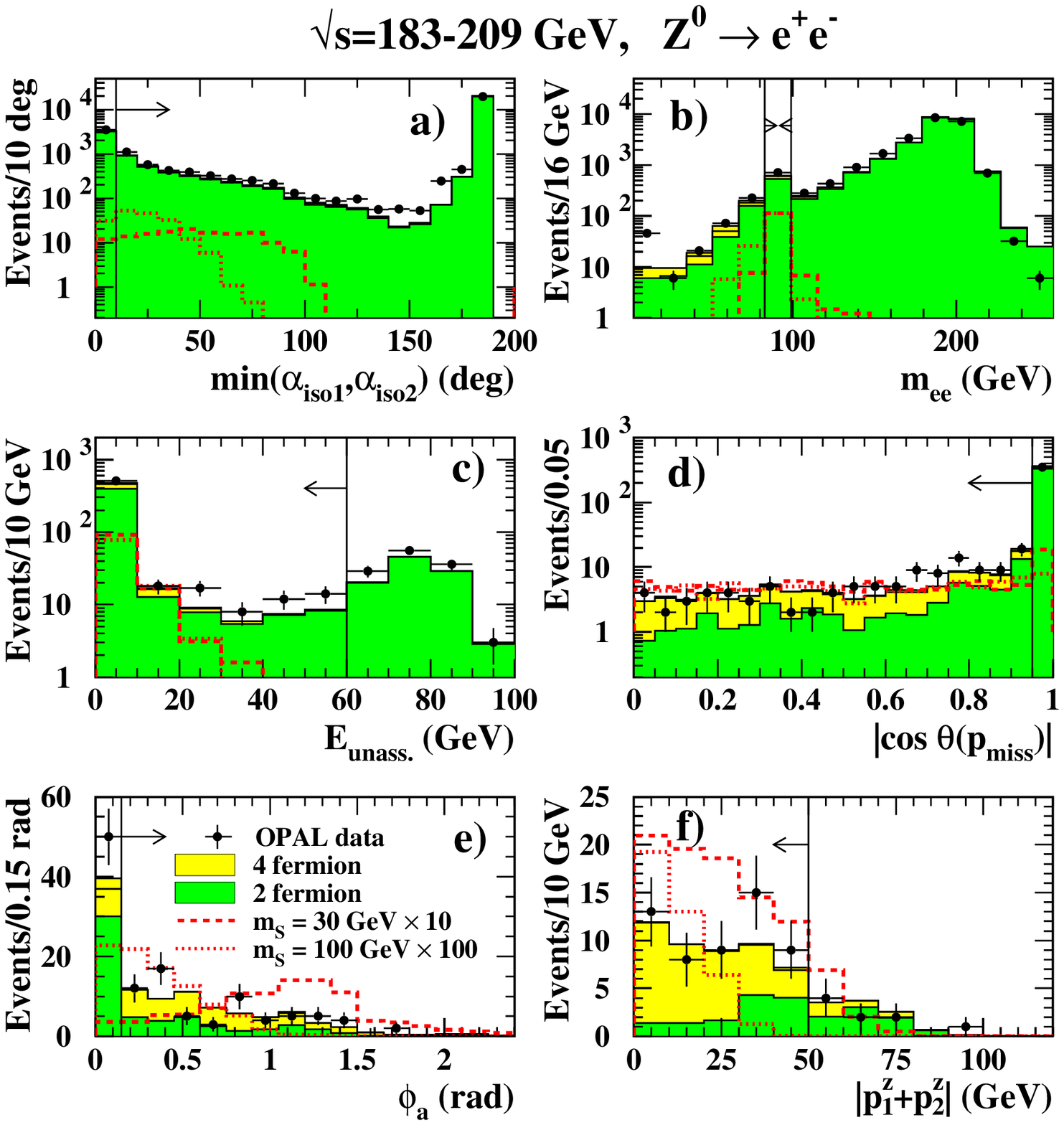}
\caption{\label{f:cutvars_eeLEP2}
  Cut variables for $\Zzero \to \ee$ at $\sqrt{s} = 183$--$209~\GeV$.
  The \klein{OPAL} data are indicated by dots with error bars
  (statistical error), the four-fermion background by the light grey
  histograms and the two-fermion background by the medium grey
  histograms. The signal distributions from a 30~GeV \Szero are
  plotted as dashed lines and those from a 100~GeV \Szero as dotted
  lines, respectively. The signal histograms are normalised to 10 and
  100 times of the Standard Model Higgs-strahlung cross section,
  respectively, and show the decays $\Szero\to\mathrm{gg}$. Each
  variable is shown with the cuts applied before the cut on this
  variable is done, respecting the order of cuts in
  Table~\ref{t:cutflow_LEP2}. The arrows indicate the accepted
  regions.
  }
\end{figure}

\begin{figure}
\centering
\includegraphics[width=16cm]{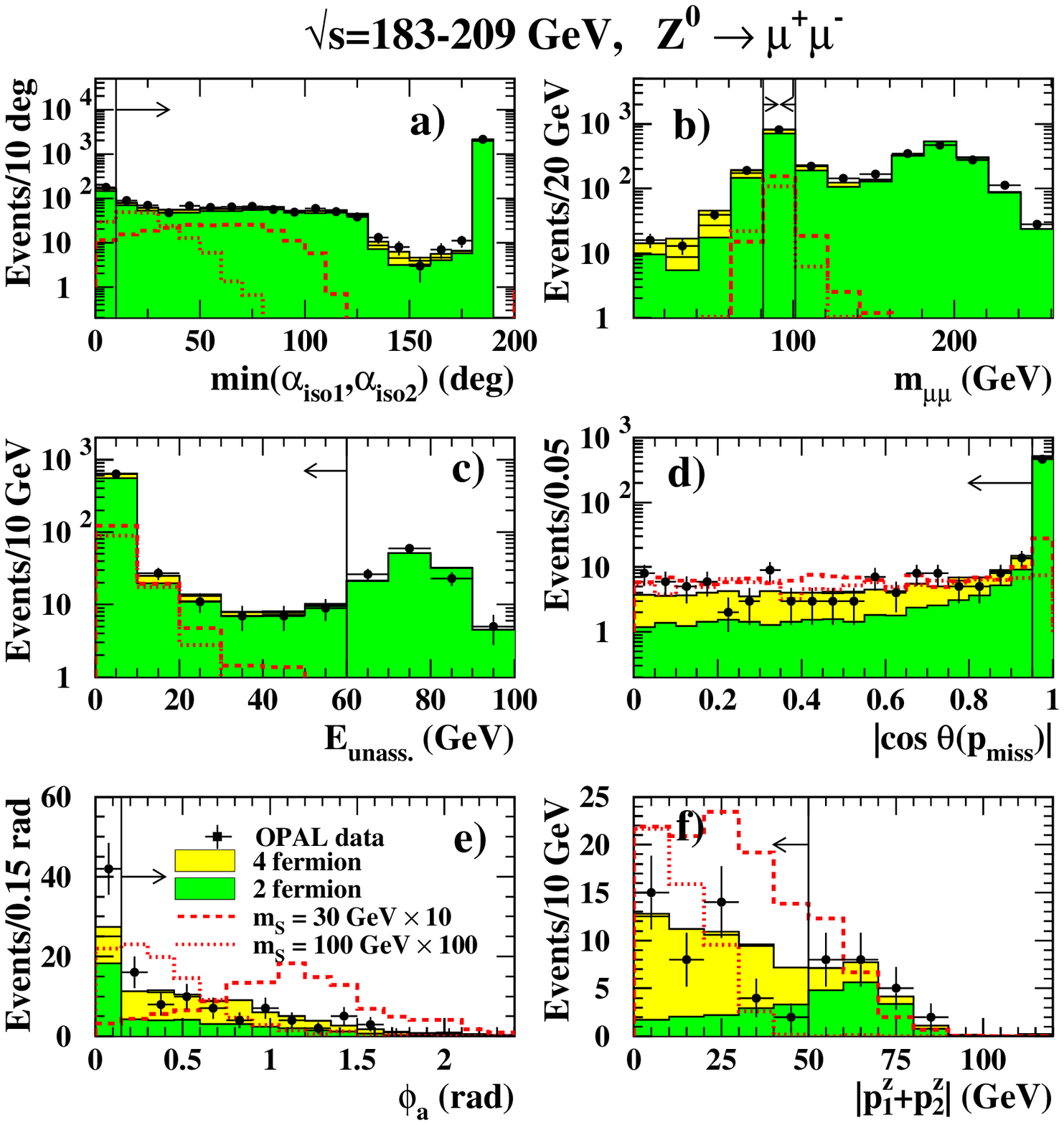}
\caption{\label{f:cutvars_mmLEP2}
  Cut variables for $\Zzero \to \mm$ at $\sqrt{s} = 183$--$209~\GeV$.
  The \klein{OPAL} data are indicated by dots with error bars
  (statistical error), the four-fermion background by the light grey
  histograms and the two-fermion background by the medium grey
  histograms. The signal distributions from a 30~GeV \Szero are
  plotted as dashed lines and those from a 100~GeV \Szero as dotted
  lines, respectively. The signal histograms are normalised to 10 and
  100 times of the Standard Model Higgs-strahlung cross section,
  respectively, and show the decays $\Szero\to\mathrm{gg}$. Each
  variable is shown with the cuts applied before the cut on this
  variable is done, respecting the order of cuts in
  Table~\ref{t:cutflow_LEP2}. The arrows indicate the accepted
  regions.
  }
\end{figure}

All cuts are listed in Table~\ref{t:cuts_lep2} and the number of
events after each cut is given in Table~\ref{t:cutflow_LEP2}.  The
distributions of the cut variables in data and Monte Carlo are shown
in Figures~\ref{f:cutvars_eeLEP2} and \ref{f:cutvars_mmLEP2} for data
taken at \sqrts = 183--209~\GeV.  A total of 54 events remain in the
data of 183--209~\GeV in the channel $\Zzero \to \ee$, with
\err{46.9}{0.6}{3.3} events expected from \sm background (the
evaluation of the systematic uncertainties is described in
section~\ref{s:ee,mm_systematics}). In the channel $\Zzero \to
\mu^+\mu^-$, 43 events remain in the data with \err{51.6}{0.3}{2.6}
expected from \sm background.  The signal efficiency is at least 24\%
in the electron channel and at least 30\% in the muon channel for
\Szero masses between 30 and 90~\GeV.
  
\begin{figure}
  \centering
  \includegraphics[width=9cm]{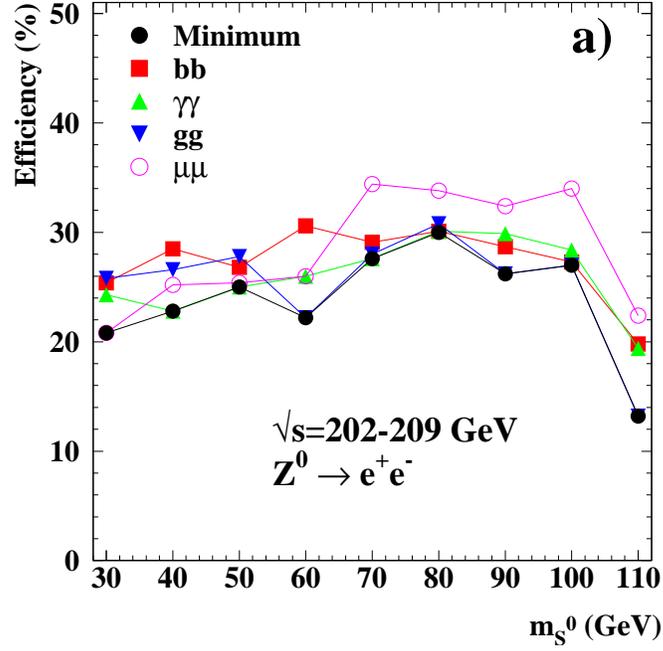}\\[4ex]
  \includegraphics[width=9cm]{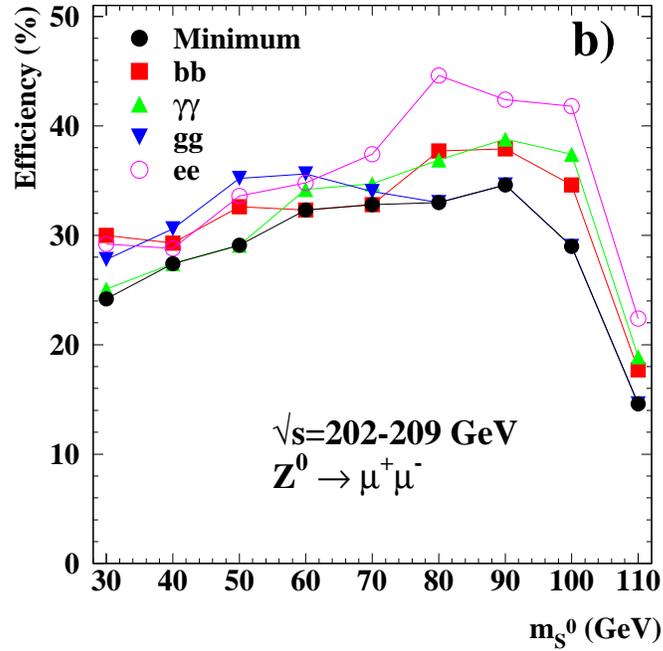}
  \caption{\label{f:eff196}
    The efficiency versus the \Szero mass at $\sqrt{s}=202$--$209$~GeV
    for a subset of decay modes of \Szero a) in the $\Zzero\to\ee$ and
    b) $\Zzero\to\mm$ channel. The minimum efficiencies which are used
    in the limits are given as well. For the other \LEP~2
    centre-of-mass energies the signal efficiencies are similar.
    }
\end{figure}

\begin{figure}
  \centering
  \includegraphics[width=15cm]{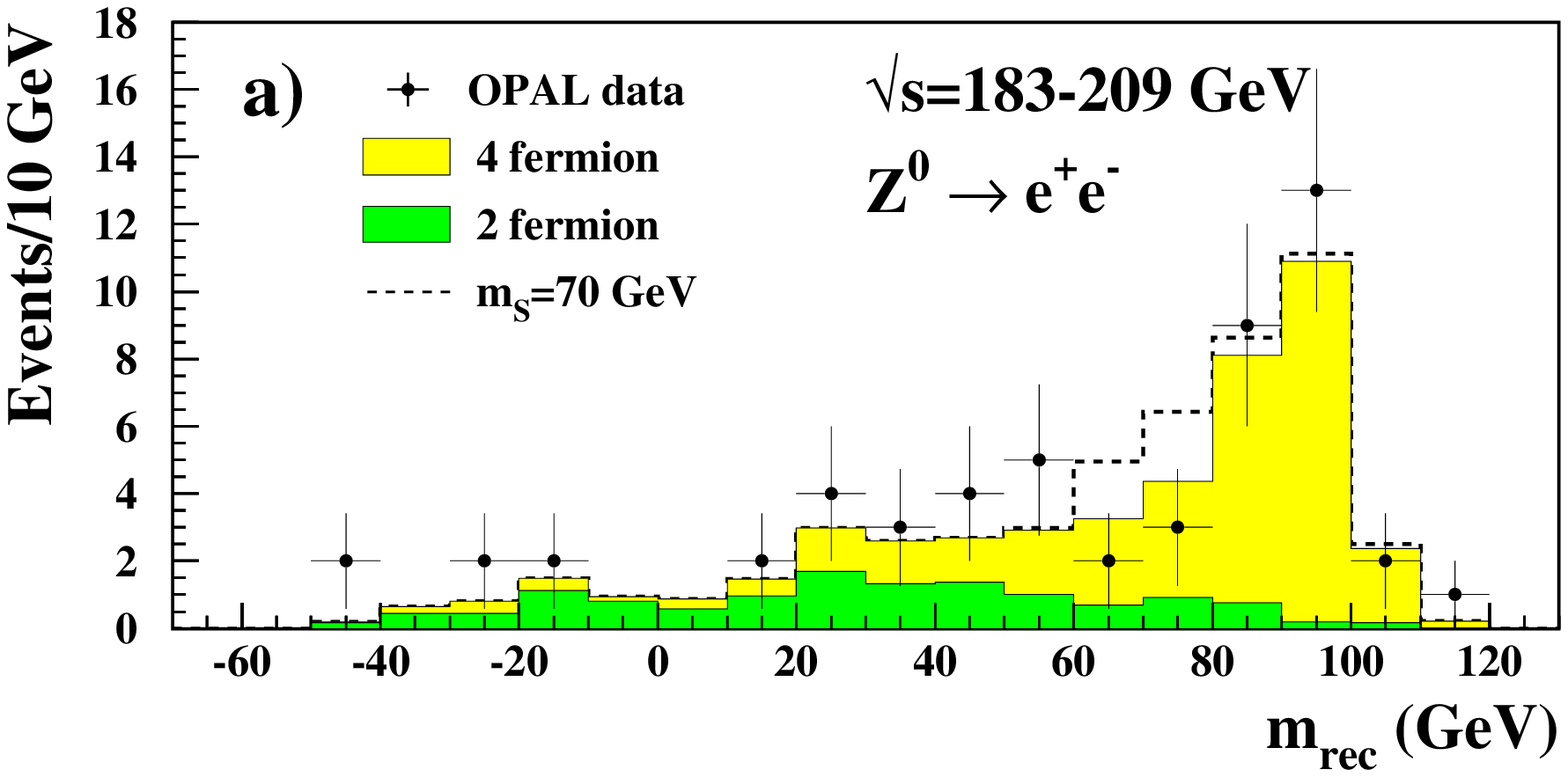}\\[5ex]
  \includegraphics[width=15cm]{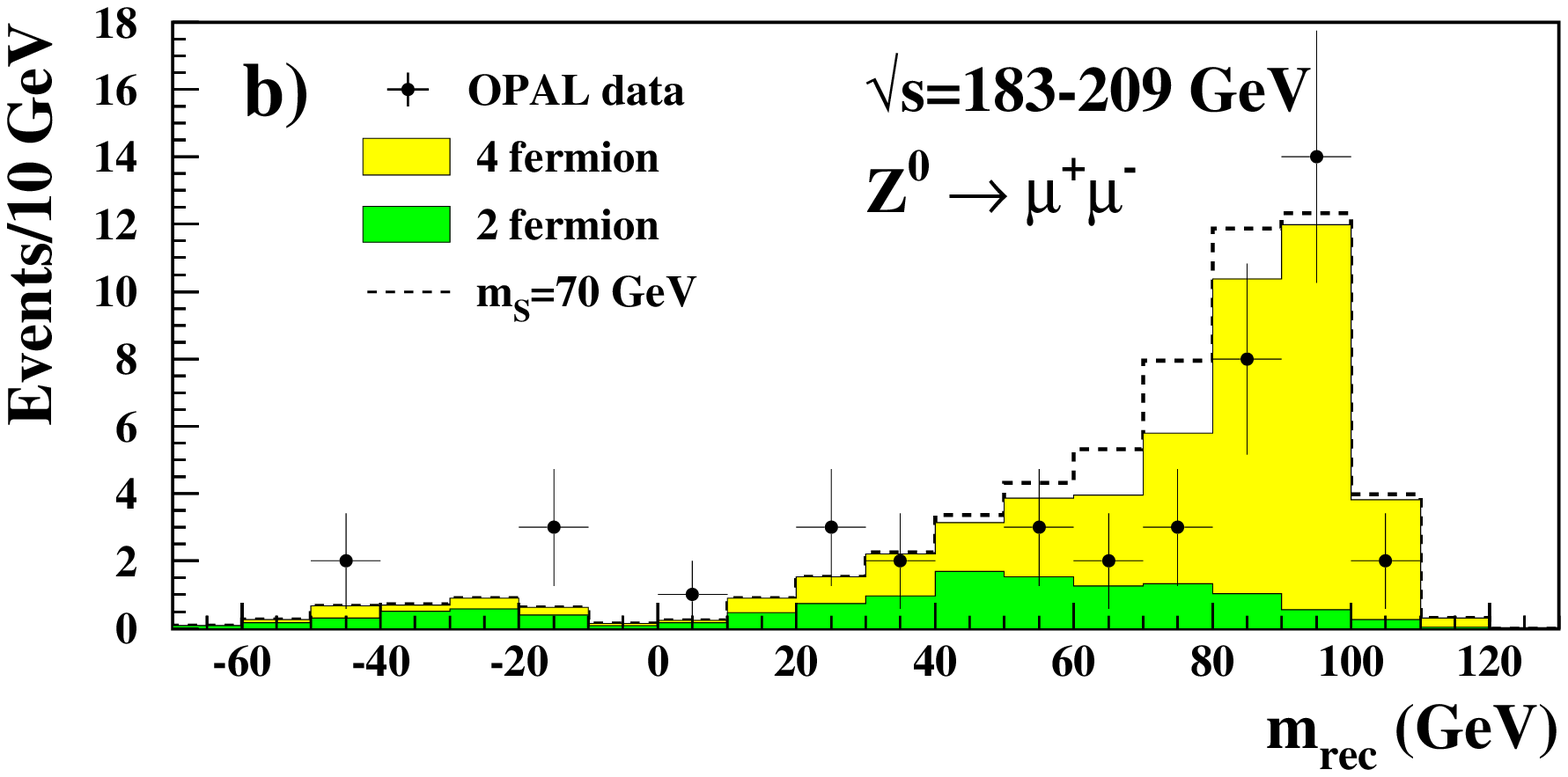}
\caption{\label{f:summass_LEP2} 
  The recoil mass spectrum from 183--209~GeV a) for the decays $\Zzero
  \to \ee$ and b) for $\Zzero \to \mm$ (lower plot). \klein{OPAL} data
  are indicated by dots with error bars (statistical error), the
  four-fermion background by the light grey histograms and the
  two-fermion background by the medium grey histograms. The dashed
  lines for the signal distributions are plotted on top of the
  background distributions with normalisation corresponding to the
  excluded cross section from the combination of both channels.
  }
\end{figure}

Figure~\ref{f:eff196} shows the efficiency versus the \Szero mass at
$\sqrt{s}=202$--$209~\GeV$ for some example decays as well as the
minimum efficiencies which are used in the limit calculation.  The
efficiencies for 183--202~\GeV have similar values for $\mS <
100~\GeV$. For the lower centre-of-mass energies the efficiency
decreases faster for higher masses due to kinematic effects, primarily
the cut on the acoplanarity angle.  Figure~\ref{f:summass_LEP2} shows
the recoil mass spectrum for both channels summed from 183--209~\GeV.
  
\subsubsection{Correction on background and signal efficiencies}
\label{s:RBC}
In all channels a correction is applied to the number of expected
background events and the signal efficiencies due to noise in the
detectors in the forward region which is not modelled by the Monte
Carlo. The correction factor is derived from the study of random beam
crossings.  The fraction of events that fail the veto on activity in
the forward region is 7.5\,\% for \LEP~1 and 3.1\,\% for \LEP~2. Since
the veto is applied only to events with less than or equal to four
tracks, the corrections on the expected background in the actual
analyses are typically only 1.8--3.5\,\%.  For the signal efficiencies
the full correction is applied to the decay channels where
appropriate.

\subsubsection{Systematic uncertainties}\label{s:ee,mm_systematics}
The systematic uncertainty of the lepton identification efficiency is
studied in a control sample of events with two collinear tracks of
which at least one is tagged as an electron or muon. The systematic
uncertainty is obtained from the difference of the identification
efficiencies for the other track between data and Monte Carlo.

The tracking systematics are studied by changing the track
resolution\footnote{$d_0$ is the distance between the vertex and the
  point of closest approach of a track to the vertex in the
  $r$--$\phi$ plane, $z_0$ is the $z$-coordinate of the track at this
  point, and $\kappa$ is its curvature.} in the Monte Carlo by a
relative fraction of 5\% in $d_0$ and $\phi$ and by 10\% in $z_0,
\kappa$ and $\cot\theta$, which corresponds to the typical difference
in the resolution of these parameters in data and Monte Carlo.  The
difference in signal and background expectation compared to the one
obtained from the unchanged track resolution is taken as the
systematic uncertainty.

The reconstruction of the energy deposition in the electromagnetic
calorimeter and the momentum in the tracking system of the lepton
candidates is investigated with the help of the mean values
$\bar{x}_{dat}$ and $\bar{x}_{MC}$ of the distributions of $p$ and $E$
from the collinear lepton pair control sample for data and Monte Carlo
expectation.  The analyses are repeated with the cuts on $p$ and $E$
being changed by the difference $|\bar{x}_{dat} - \bar{x}_{MC}|$. The
deviations in the number of expected events compared to the original
cuts are taken as the systematic uncertainties.

The uncertainty from the lepton isolation angle \aiso is studied in
different ways for the \LEP~1 and \LEP~2 analysis. In the \LEP~1
selection a control sample of hadronic events is selected.  Random
directions are then chosen in the event, and the angles \aiso of the
vectors pointing to these directions are determined. The cut on \aiso
is varied by the difference of the mean of the data and the \MC
distributions of the control sample. In the \LEP~2 selection the
uncertainty is obtained in a similar way but from the isolation of the
lepton in W$^+$W$^-\to\mathrm{q\bar{q}}\ell\nu$ events.

Correct modelling of photon radiation and conversions is a crucial
ingredient of the decay-mode independent searches. For \LEP~1, the
effect of the description of photon radiation in the Monte Carlo is
estimated from the difference in the number of events between data and
background expectation after removing the photon and conversion veto.
At least one identified photon or conversion is required for the
tested events. In the muon channel at \LEP~2 energies two different
Monte Carlo generators are used for the two-fermion background, and
the difference between the background prediction of the two generators
is taken as the systematic uncertainty of the photon modelling. For
the electron channel only one generator is available. Here, the
uncertainty is determined from the comparison of the number of events
in the data and Monte Carlo sets in a side band of the distribution of
the lepton pair invariant mass where no signal is expected.  This test
is dominated by the statistical uncertainties of the side-band
sub-sample.

In the analysis the four-fermion Monte Carlo samples are reweighted to
account for low mass resonances (e.g. $\rho, \omega, \phi, J/\psi$)
and the running of $\alpha_{\mathrm{em}}(q^2)$. The uncertainty from
this reweighting is assessed to be 50\% of the change of the expected
background after switching off the reweighting.

All uncertainties for a particular centre-of-mass energy are assumed
to be uncorrelated and the individual contributions are added in
quadrature for the total systematic uncertainty. The dominant
systematic uncertainties in the \LEP~1 background expectation come
from the description of photon radiation and photon conversions in the
Monte Carlo as well as from the uncertainty of the four-fermion
cross section. The precision of the predicted signal efficiency is
mainly limited by the description of the lepton isolation.

In the \LEP~2 selection the modelling of the radiative returns has a
large impact on the total systematic uncertainty, both in the electron
and the muon channel. In the electron channel the uncertainty from the
isolation angle criterion and in the muon channel the uncertainty of
the muon identification efficiency are also significant.

For the \LEP~2 data, the evaluation of the systematic uncertainties at
each single centre-of-mass energy is limited by Monte Carlo
statistics. Therefore they are investigated for the total set of
Monte Carlo samples with $\sqrt{s} = 183$--$209~\GeV$.

The numbers of expected background events for the \LEP~1 and \LEP~2
analyses, broken down by the different centre-of-mass energies, are
listed in Table~\ref{t:d_b_error} for the channels $\Zzero \to \ee$
and $\Zzero \to \mm$. The numbers include systematic errors discussed
above and the statistical error from the limited Monte Carlo samples.
Also the number of expected events from a 30~\GeV \SM Higgs boson is
shown. A detailed overview of the different systematic uncertainties
is given in Table~\ref{tab:sys_err}.

\subsection{Event selection for \boldmath\ee$\to$ S$^0$Z$^0\to$
  $n\cdot\gamma\,\nn$, $\ee\nn$\unboldmath \label{s:neutralchannel}}
In this section a search for \ee$\to$ S$^0$Z$^0\to$
$n\cdot\gamma\,\nn$ or $\ee\nn$ (the latter 
for $m_{\Szero} < 500~\MeV$) at $\sqrt{s} = 91.2~\GeV$ is described.
The removal of radiative backgrounds in the selection described in
section~\ref{s:lep1-selection} rejects the \Szero decays into photons
(due to the photon veto) and, in the mass region $m_{\Szero} <
500~\MeV$, also the decays into electrons (due to the conversion
veto). The selection for S$^0$Z$^0\to$ S$^0$\nn\unboldmath is included
to recuperate the sensitivity to the photon and electron decay modes
and therefore to remain decay-mode independent.

\subsubsection{Event selection}\label{s:misse_event_selection}
In signal events the \Szero is radiated off the \Zzero with some
kinetic energy and a certain amount of transverse momentum. Therefore
the total visible energy $E_{\mathrm{vis}}$ in the electromagnetic
calorimeter is required to exceed 12 GeV and the transverse momentum
$p_T$ reconstructed from the event is required to exceed 6 GeV. Since
the \Zzero decays into neutrinos which carry energy out of the
detector, the total amount of visible energy is reduced. The selection
requires $E_{\mathrm{vis}} < 60~\GeV$.
  
Several cuts are applied to reduce the background from processes with
topologies different from the signal.  The selection allows only
events with zero ($n\cdot \gamma\,\nn$-channel) or two
(\ee\nn-channel) identified electrons, using the same electron
identification routines as described in section~\ref{s:LEP1_cl}.  The
next selection cuts are intended to further reduce background from
cosmic rays or beam halo particles.  Events triggered by cosmic rays
or beam halo particles are characterised by extended clusters in the
electromagnetic calorimeter, hits in the hadron calorimeter and muon
chambers and a signal from the time-of-flight counter that shows a
significant discrepancy from its expected value.  We therefore require
no hit in the muon chambers and at most two hits in the hadron
calorimeter. No more than one cluster with an energy deposition larger
than 2~\GeV is allowed in the hadron calorimeter.  The number of lead
glass blocks in each cluster in the electromagnetic calorimeter 
must be less than 15. The difference between the measured
time of flight and the expected time for a particle coming from the
interaction point is required to be less than 2~ns.
  
The remaining background is mostly from $\ee \to \nu\bar{\nu}\gamma$
events, where the photon is usually emitted at small angles to the
beam axis.  A hard cut on the angular distribution of clusters in the
electromagnetic calorimeter and the forward calorimeters (polar angle
region 47--200~mrad) is applied.  For this purpose the polar angle of
the energy vector \Evec is defined as
\[
  \theta_{\Evec} = \sum_{i=1,n} \frac{E_i\cdot\theta_i}{E_{\mathrm{vis}}}.
\]
The sum runs over all clusters $i$ (with polar angle $\theta_i$).
Cuts applied on the energy vector and the individual clusters are
$|\cos\theta_{\Evec}| < 0.65$ and $|\cos\theta_i| < 0.9$. The energy
in the forward calorimeters must be less than 2~\GeV.

Events with two electrons must satisfy some additional requirements.
The tracks must be identified as electrons with opposite charge. The
angle $\Delta\phi$ between the tracks must be less than \degree{10},
the invariant mass $m_{\ell\ell}$ must be less than 2~\GeV, and a
transverse momentum $p_T > 7~\GeV$ of the event is required. Events
are rejected if there are any additional clusters other than those
associated with one of the two tracks.

A correction due to random detector occupancy is applied as described
in section~\ref{s:RBC}. The full correction of $-7.5\,\%$ is used
since the forward detector veto applies to all events.

\begin{figure}

\centering
\includegraphics[width=16cm]{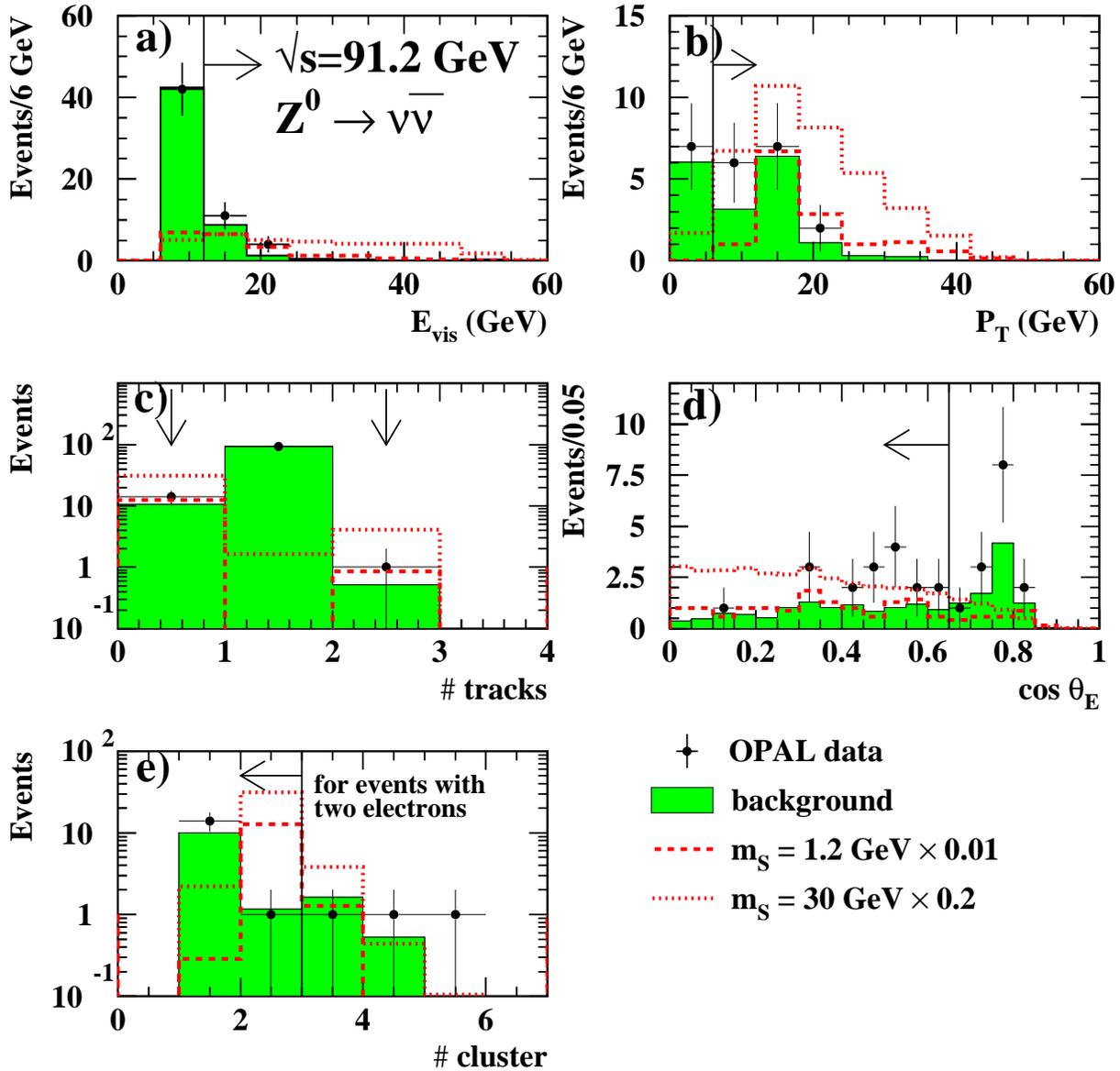}
\caption{\label{f:cutvars_nnLEP1}
  Some of the cut variables for $\Zzero \to \nu\bar{\nu}$ at
  $\sqrt{s}$ = 91.2~GeV. The last plot (number of clusters) is one of
  the additional cut variables which are used in events with two
  electrons. The \klein{OPAL} data are indicated by dots with error
  bars (statistical error), and the total background by the grey
  histograms.  The distributions from a 1.2~\GeV and a 30~\GeV signal
  are plotted as dashed and dotted lines, respectively. The signal
  histograms are normalised corresponding to 0.01 times and 0.2 times
  the \SM Higgs-strahlung cross section and show the decay channel
  $\Szero \to \gamma\gamma$. Each variable is shown with all cuts but
  the cut on the variable itself. The arrows indicate the accepted
  regions.  
  }
\end{figure}

After all cuts 15 events are selected from the data with a background
prediction of $11.3 \pm 1.1({\mathrm{stat.}})\pm
0.2({\mathrm{syst.}})$, where the uncertainties are evaluated as
described below.  Figure~\ref{f:cutvars_nnLEP1} shows the distribution
of the cut variables in data and Monte Carlo.
 
There is no statistically significant excess in this channel, and the
shape of the distributions of the cut variables in data and \MC are in
good agreement. Furthermore the product of the signal efficiency and
the $\Zzero\to\nn$ decay branching ratio is substantially higher than
for the \Szero$\ell^+\ell^-$-channels, and the predicted background is
much less.  Hence, the channel $\Zzero \to \nn$ has much higher
sensitivity than the electron and the muon channel. It does not
contribute to the actual limits, provided that the systematic
uncertainties are not much larger than in the other channels. For a
conservative limit calculation only the channels with lowest
sensitivity are used.

The search channel described in this section recovers sensitivity to
the decay modes $\Szero\to n\cdot\gamma$ and at low masses \mS to the
decay $\Szero\to\ee$ to which the analysis described in
section~\ref{s:lep1-selection} had no sensitivity.  However, the
requirement $E_{\mathrm{vis}} > 12~\GeV$ in this channel can lead to
an insensitivity to decays $\Szero \to \chi^0_2\chi^0_1 \to
\gamma\chi^0_1\chi^0_1$ for certain combinations of $\mS,
m_{\chi^0_2}$ and $m_{\chi^0_1}$ for the whole \LEP~1 analysis.

\subsubsection{Systematic uncertainties}
Uncertainties in this channel predominantly come from the energy
calibration of the electromagnetic calorimeters. In
reference~\cite{c:calibration} it is shown that an electromagnetic
cluster has a calibration uncertainty of 25~\MeV. Since the number of
clusters in the electromagnetic calorimeter for selected data and \mc
events is less than five, the deviation of the expected number of
background events and the signal efficiencies after shifting the cuts
on the visible energy $E_{\mathrm{vis}}$ and the transverse momentum
$p_T$ by four times 25~\MeV is used as a systematic uncertainty.  The
deviation is found to be 1.4\,\%.  From the same
reference~\cite{c:calibration} we take the systematic uncertainties on
the time-of-flight signal (0.5\,\%). Other sources for systematic
uncertainties are the luminosity (0.5\,\%), the limited Monte Carlo
statistics (10.0\,\% for background and 2.2\,\% for the signal) and,
for events where the \Szero decays into electrons, the uncertainty on
the electron identification efficiency (0.8\,\%).  Summing these
individual sources up in quadrature, estimates of the total
uncertainty in the background of 10.0\,\%(stat.)$+$1.8\,\%(syst.) and
in the signal of 2.2\,\%(stat.)$+$1.6\,\%(syst.) are obtained.
Given the expected number of signal and background events, this is
much less than the level where the channel starts contributing to the
limit calculation.

%
%
\section{Results}\label{s:results}
The results of the decay-mode independent searches are summarised in
Table~\ref{t:d_b_error}, which compares the numbers of observed
candidates with the background expectations.  The total number of
observed candidates from all channels combined is 208, while the \SM
background expectation amounts to $207.3\pm 4.1(\mathrm{stat.}) \pm
11.1(\mathrm{syst.})$.  For each individual search channel there is
good agreement between the expected background events and observed
candidates.  As no significant excess over the expected background is
observed in the data, limits on the cross section for the Bjorken
process $\mathrm{e}^+\mathrm{e}^-\to$ \Szero{}\Zzero\ are calculated.
  
The limits are presented in terms of a scale factor \sq, which relates
the cross section for \Szero{}\Zzero to the \SM\ one for the
Higgs-strahlung process $\mathrm{e}^+\mathrm{e}^-\to$ \HSM{}\Zzero as
defined in Equation~\ref{e:sq_def}. The 95\% CL upper bound on $\sq$
is obtained from a test statistic for the signal hypothesis, by using
the weighted event-counting method described in~\cite{c:mssmpaper172}:
In each search channel, given by the different centre-of-mass energies
and the $\Zzero$ decay modes considered, the observed recoil mass
spectrum is compared to the spectra of the background and the signal.
The latter is normalised to $\varepsilon \cdot \mathrm{BR} \cdot
{\cal{L}} \cdot \sq \cdot \sigma_{\mathrm{\Hzero\Zzero}}$, where
$\varepsilon$ is the minimum signal detection efficiency out of all
tested decay modes, BR is the branching ratio of the $\Zzero$ decay
mode considered in this channel and $\cal{L}$ is the integrated
luminosity recorded for that channel. The efficiencies for arbitrary
\Szero masses are interpolated from the efficiencies at masses for
which Monte Carlo samples were generated. Every event in each of these
mass spectra and each search channel is given a weight depending on
its expected ratio of signal over background, $s/b$, at the given
recoil mass. For every assumed signal \Szero mass these weights are a
function of the signal cross section, which is taken to be $\sq$ times
the Standard Model Higgs cross section for the same \Szero mass.
Finally, from the sum of weights for the observed number of events, an
upper limit $k^{95}$ for the scale factor is determined at the 95\%
confidence level.

The systematic uncertainties on the background expectations and signal
selection efficiencies are incorporated using the method described
in~\cite{c:cousins}.
  
The limits are given for three different scenarios:
\begin{enumerate}
\item Production of a single new scalar {\boldmath S$^0$\unboldmath}.
\item The Uniform Higgs scenario.
\item The Stealthy Higgs scenario.
\end{enumerate}

\subsection{Production of a single new scalar \boldmath S$^0$\unboldmath}
\label{general}
In the most general interpretation of our results, a cross-section
limit is set on the production of a new neutral scalar boson \Szero in
association with a \Zzero boson. To calculate the limit we use the
mass distributions of which the sums are shown in
Figure~\ref{f:summass_LEP1} and \ref{f:summass_LEP2} for \OPAL data,
the expected background and the signal.

\begin{figure}
  \centering
  \includegraphics[width=0.85\textwidth]{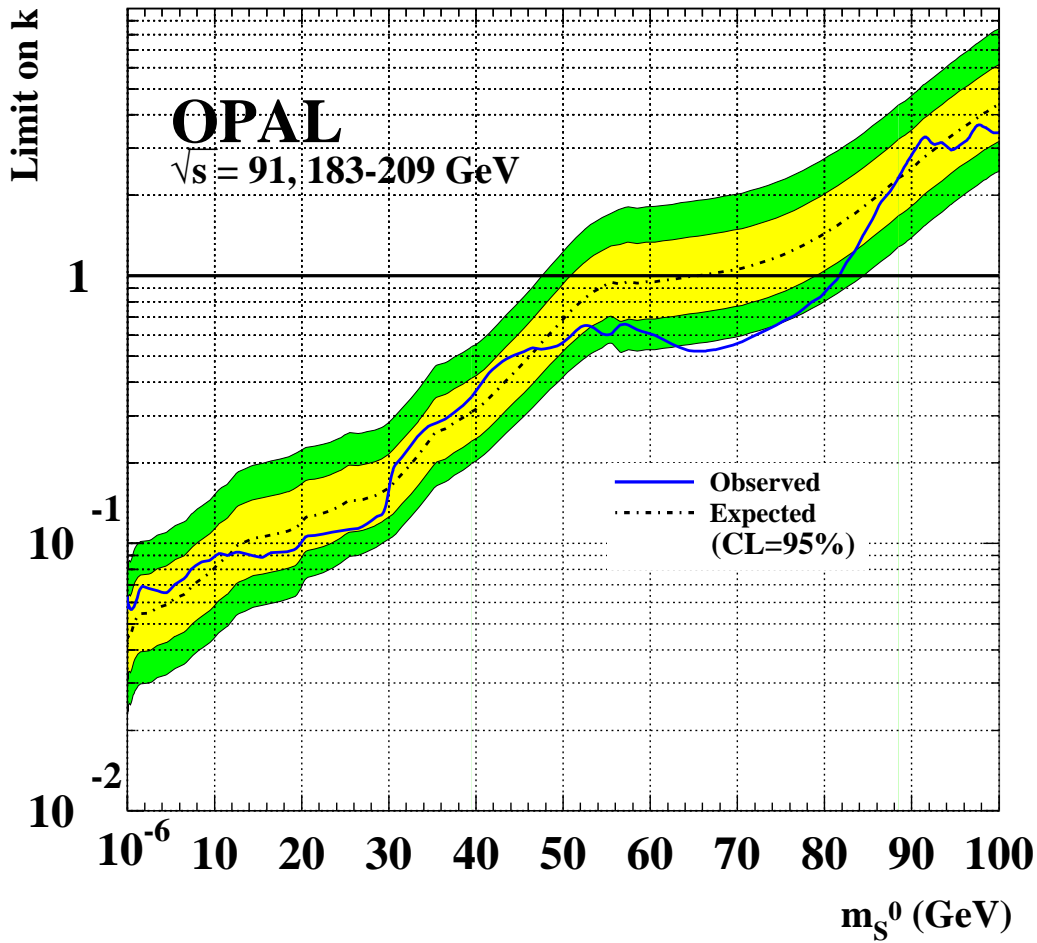}
\caption{
  The upper limit on the scale factor \sq on the cross section for the
  production of a new scalar boson in the Higgs-strahlung-process
  (solid line). The dot-dashed line represents the expected median for
  background-only experiments.  Both limits are calculated at the
  95\,\% confidence level.  The dark (light) shaded bands indicate the
  68\% (95\%) probability intervals centred on the median expected
  values.  For masses $\mS \lesssim 1~\GeV$ the limits are constant.
  The lowest signal mass tested is $10^{-6}~\GeV$.  
  }
\label{f:di_limits}
\end{figure}

In Figure~\ref{f:di_limits} we present the limits obtained for scalar
masses down to the lowest generated signal mass of 1~\keV. They are
valid for the decays of the \Szero into hadrons, leptons, photons and
invisible particles (which may decay inside the detector) as well as
for the case in which the \Szero has a sufficiently long lifetime to
escape the detector without interacting or decaying.  A decay of the
\Szero into invisible particles plus photons, however, can lead to a
reduced sensitivity in the mass region where the sensitivity of the
analyses is dominated by the \LEP~1 data (see
section~\ref{s:misse_event_selection}).  The observed limits are given
by the solid line, while the expected sensitivity, determined from a
large number of Monte Carlo experiments with only background, is
indicated by the dotted line. The shaded bands indicate the one and
two sigma deviations from the expected sensitivity. Values of $\sq >
0.1 $ are excluded for values of \mS below 19~\GeV, whereas $\sq > 1 $
is excluded from the data for \mS up to 81~\GeV, independent of the
decay modes of the \Szero\ boson. This means that the existence of a
Higgs boson produced at the SM rate can be excluded up to this mass
even from decay-mode independent searches.  For masses of the new
scalar particle well below the width of the \Zzero, \emph{i.e.} $\mS
\lesssim 1~\GeV$, the obtained limits remain constant at the level of
$\sq^{95}_{\mathrm{obs.}} = 0.067$, and $\sq^{95}_{\mathrm{exp.}} =
0.051$.

The discrepancy between the expected and the observed limits is within
one standard deviation for masses below 52~\GeV and for masses above
82~\GeV. The deviation of about two sigma in the mass range
52--82~\GeV is due to a deficit of selected data events in the recoil
mass spectrum of both the electron and muon channels.

\subsection{Limits on signal mass continua}
\subsubsection{The Uniform Higgs scenario\label{s:uniform-higgs}}
We simulated signal spectra for the Uniform Higgs scenario for
$\Ktilde = \mathrm{constant}$ over the interval [\mA, \mB{}] and zero
elsewhere.  Both the lower mass bound \mA and the upper bound \mB are
varied between 1~\GeV and 350~\GeV (with the constraint $\mA\le\mB$).
In a similar way to the previous section we get an upper limit on the
integral in Equation~\ref{eq:sumrule1}.
  
\begin{figure}
  \centering
  \includegraphics[width=0.85\textwidth]{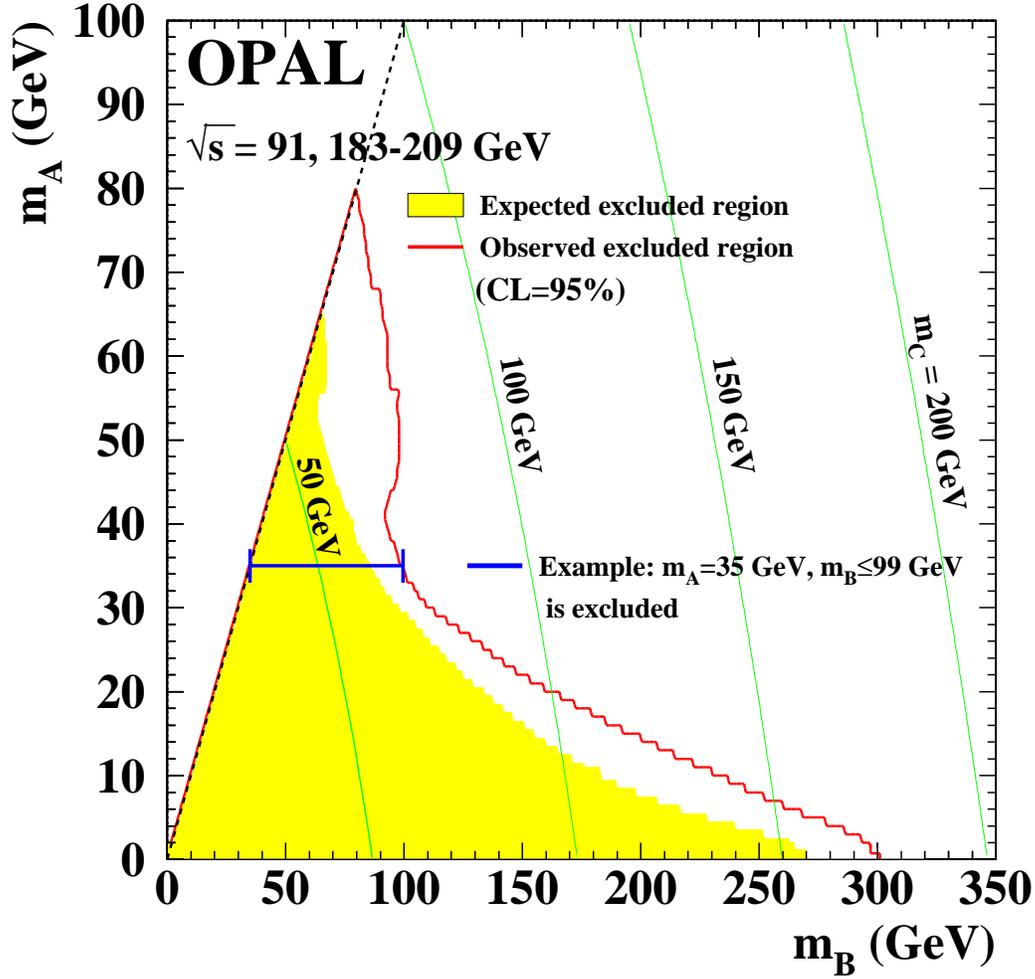}
\vspace*{-2ex}

\caption{
  Exclusion limits for the Uniform Higgs scenario at the 95\,\%
  confidence level. All mass intervals (\mA,\,\mB) within the area
  bordered by the dark line are excluded from the data. The shaded
  area marks the mass points which are expected to be excluded if
  there were only background.  The light grey curves indicate isolines
  for several values of \mC. All intervals $(\mA,\mB)$ to the right of
  each isoline are theoretically disallowed from
  Equation~\ref{eq:sumrule2}. By definition, only intervals
  $(\mA,\mB)$ right to the dashed diagonal line are valid, \emph{i.e.}
  $\mA \le \mB$.
  \label{fig:excluded_continuum_higgs}
  }
\end{figure}

Figure~\ref{fig:excluded_continuum_higgs} shows the mass points
(\mA,\mB) for which the obtained 95\,\%~CL limit on
$\int\mathrm{d}m\,\Ktilde$ is less than one.  These are the signal
mass ranges $\mA \le \mhi \le \mB$ which can be excluded assuming a
constant \Ktilde.

If $\mA=\mB$, then the signal spectrum reduces to the mass
distribution of a single boson. Excluded points on the diagonal
$\mA=\mB$ are therefore the same masses as in Figure~\ref{f:di_limits}
for which $k < 1$.  The horizontal line illustrates an example for
excluded mass ranges: The line starts on the diagonal at $\mA = \mB =
35~\GeV$ and ends at $\mB = 99~\GeV$. This value of \mB is the highest
upper mass bound which can be excluded for this value of \mA.  All
mass ranges with an upper bound \mB below 99~\GeV are also excluded
for $\mA = 35~\GeV$.  The highest excluded value of \mB ($\mB =
301~\GeV)$ is achieved for $\mA$ set to 0~\GeV.

\begin{figure}
  \centering
  \includegraphics[width=0.85\textwidth]{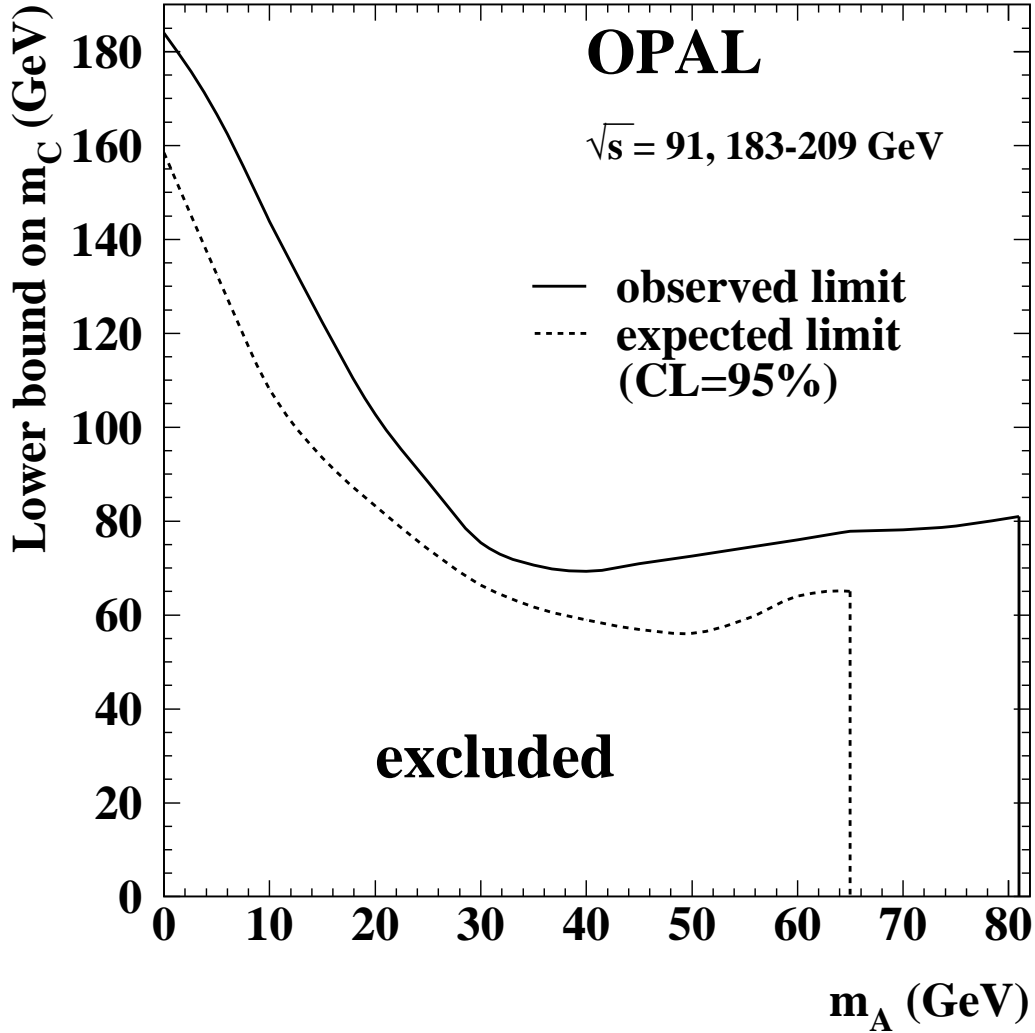}
\vspace*{-2ex}

\caption{
  Exclusion limits on the perturbative mass scale \mC for constant
  \Ktilde. The solid line represents the limits obtained from the
  data, and the dotted line shows the expected limit if there were
  only background. Values for \mC below the lines are excluded by this
  analysis at the 95\,\% confidence level.  
  }
  \label{fig:excluded_mC}
\end{figure}

Using the two sum rules from section~\ref{worst-case}, lower limits on
the perturbative mass scale \mC can be derived. For each excluded
value of \mA we take the highest excluded value of \mB and determine
the lower bound of \mC according to Equation~\ref{eq:sumrule2}.  The
excluded mass ranges for \mC, assuming a constant \Ktilde, are shown
in Figure~\ref{fig:excluded_mC}.

\subsubsection{Bin-by-bin limits}
The limits presented in section~\ref{s:uniform-higgs} are specific to
the case where the coupling density is constant in the interval [\mA,
\mB{}] and zero elsewhere. The data can also be used to exclude other
forms of $\Ktilde(m)$. To provide practical information for such
tests, we have measured $\Ktilde(m)$ in mass bins with a width
comparable to the experimental mass resolution.  The typical
resolution of the recoil mass in the \LEP~1 analysis varies between 1
and 5~\GeV in the mass region between 10 and 55~\GeV.  In the \LEP~2
analysis the width is between 3 and 15~\GeV for recoil masses between
20 and 100~\GeV.  The width gets smaller at higher recoil masses.  The
results of the measurement of $\Ktilde(m)$ are given in
Table~\ref{t:bin_klim} together with the corresponding statistical and
systematic uncertainties.
  
From these measured numbers of $\Ktilde(\mhwo)$, one can obtain upper
limits on the integral \mbox{$\int\mathrm{d}\mhwo \, \Ktilde(\mhwo)$}
for any assumed shape of $\Ktilde(\mhwo)$ using a simple $\chi^2$
fitting procedure.  To account for mass resolution effects, we provide
a correction matrix (Table~\ref{t:correction_matrix}). To test a
certain theory with a distribution of $\Ktilde(\mhwo)$ values in the
10 measured bins from Table~\ref{t:bin_klim}, written as a vector
$\vec{\kappa} = \left(\tilde{K}_{1},\ldots,\tilde{K}_{10} \right)$,
the corrected vector $\vec{\kappa}' = \hat{C}\vec{\kappa}$ can then be
fitted to the measured values.  In the fit the systematic
uncertainties, which are small compared to the statistical errors, can
be assumed to be fully correlated bin-by-bin.

\subsubsection{The Stealthy Higgs scenario}
\label{s:stealthy}
To set limits on the Stealthy Higgs scenario we have simulated the
spectrum of a Higgs boson with a width according to
Equation~\ref{eq:higgs_width} and Ref.~\cite{c:stealthy_higgs}.

\begin{figure}
  \centering
  \includegraphics[width=0.85\textwidth]{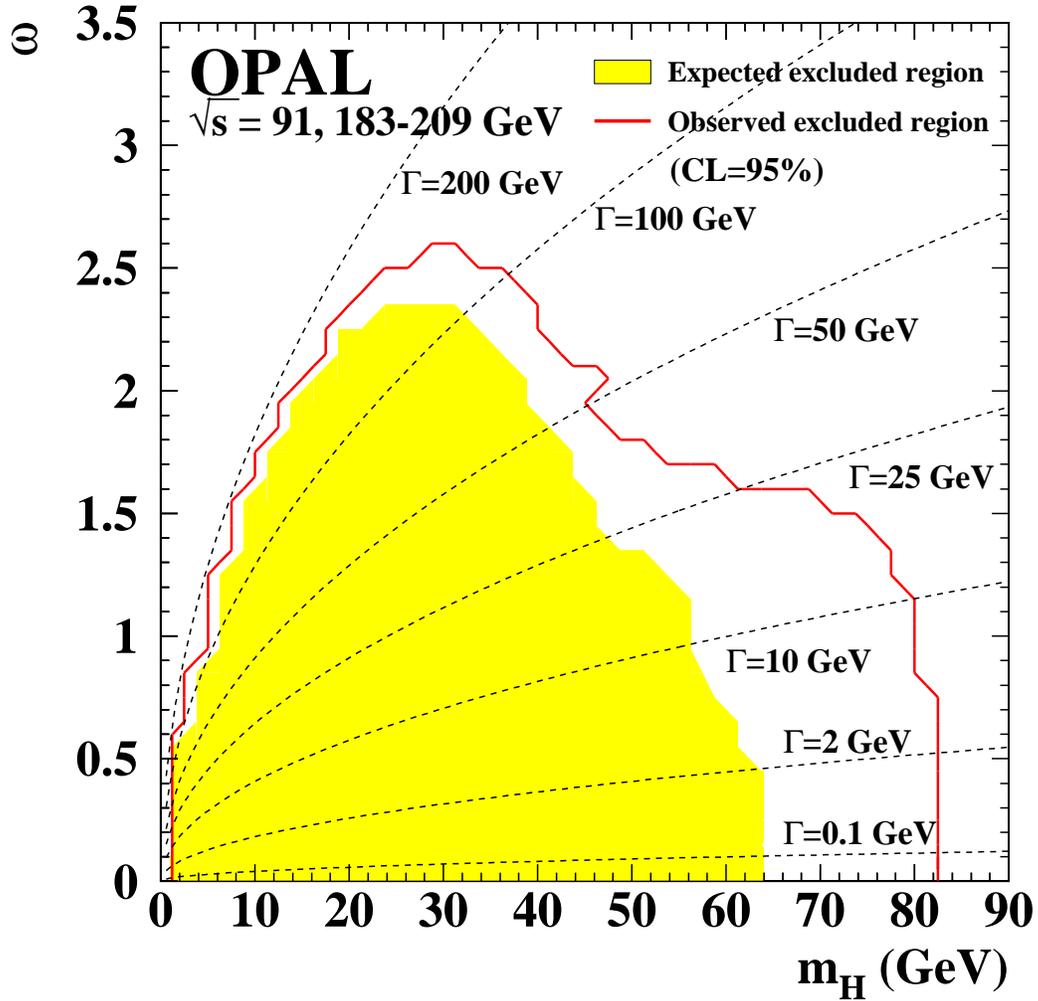}
\caption{
  Excluded parameter regions for the simplified Stealthy Higgs
  scenario at the 95\,\% confidence level. The solid line marks the
  region which is excluded from the data. The shaded area marks the
  region which would be excluded if the data corresponded exactly to
  the background-only prediction.  The dashed lines indicate the Higgs
  width depending on $m_{\mathrm{H}}$ and $\omega$.
  \label{fig:excluded_hidden_higgs}
  }
\end{figure}

The excluded regions in the $\omega$-$m_{\mathrm{H}}$ parameter space
are shown in Figure~\ref{fig:excluded_hidden_higgs}. To illustrate the
Higgs width according to Equation~\ref{eq:higgs_width}, for a given
mass $m_{\mathrm{H}}$ and coupling $\omega$ `isolines' for some sample
widths are added to the plot. The vertical edge in the exclusion
contour at $m_{\mathrm{H}}=81~(62)~\GeV$ in the observed (expected)
limits reflects the detector mass resolution in \deltam: For a fixed
mass $m_{\mathrm{H}}$ the exclusion power is the same for all
couplings $\omega$ that yield $\Gamma_{\mathrm{H}}\lesssim\deltam$,
and the limits for $\omega\to 0$ reproduce the limits for a single
narrow \Szero in Figure~\ref{f:di_limits}.  The maximal excluded
region of the coupling $\omega$ is achieved for masses around 30~\GeV,
where $\omega$ can be excluded up to $\omega = 2.7$.  For lower masses
the sensitivity drops due to the rapidly increasing width of the Higgs
boson, and for higher masses due to the decreasing signal
cross section.

\section{Conclusions}
Searches for new neutral scalar bosons \Szero decaying to hadrons of
any flavour, to leptons, photons invisible particles and other modes
have been performed based on the data collected at $\sqrt{s}$ = \mZ
and 183 to 209~\GeV by studying the recoil mass spectrum of
$\Zzero\to\ee, \mu^+ \mu^-$ in \Szero{}\Zzero production and the
channel where the \Zzero decays into $\nn$ and the \Szero into photons
or \ee. No significant excess of candidates in the data over the
expected Standard Model background has been observed.  Therefore upper
limits on the production cross section for associated production of
\Szero and \Zzero, with arbitrary \Szero decay modes, were set at the
95\,\% confidence level.  Upper limits in units of the \SM
Higgs-strahlung cross section of $\sq < 0.1$ for $1~\keV < \mS<
19~\GeV$ and $\sq < 1$ for $\mS < 81~\GeV$ were obtained.  In further
interpretations, limits on broad continuous signal mass shapes to
which previous analyses at \LEP had no or only little sensitivity were
set for the first time.  Two general scenarios in the Higgs sector
were investigated: A uniform scenario, when the signal arises from
many unresolved Higgs bosons, and a Stealthy Higgs model, when the
Higgs resonance width is large due to large Higgs-phion couplings.

\section*{Acknowledgements}
We gratefully thank J.J.\,van\,der\,Bij and T.\,Binoth for valuable
discussions concerning the Stealthy Higgs scenario.
\bigskip

\noindent 
We particularly wish to thank the SL Division for the efficient
operation of the LEP accelerator at all energies and for their close
cooperation with our experimental group. In addition to the support
staff at our own institutions we are pleased to acknowledge the  \\
Department of Energy, USA, \\
National Science Foundation, USA, \\
Particle Physics and Astronomy Research Council, UK, \\
Natural Sciences and Engineering Research Council, Canada, \\
Israel Science Foundation, administered by the Israel
Academy of Science and Humanities, \\
Benoziyo Center for High Energy Physics,\\
Japanese Ministry of Education, Culture, Sports, Science and
Technology (MEXT) and a grant under the MEXT International
Science Research Program,\\
Japanese Society for the Promotion of Science (JSPS),\\
German Israeli Bi-national Science Foundation (GIF), \\
Bundesministerium f\"ur Bildung und Forschung, Germany, \\
National Research Council of Canada, \\
Hungarian Foundation for Scientific Research, OTKA T-029328,
and T-038240,\\
Fund for Scientific Research, Flanders, F.W.O.-Vlaanderen, Belgium.\\

\clearpage

%
%

%
%
\clearpage
\begin{table}
  \small
  \begin{center}
  \begin{tabular}{|c|c|rc|}
    \hline
    $\sqrt{s}$ (GeV) &  year        &  \multicolumn{2}{r|}{integrated luminosity (pb$^{-1}$)} \\
    \hline
    91.2         &  1989--95    &\hspace*{15mm} 115.4      &    \\
    183          &  1997        &                56.1      &    \\
    189          &  1998        &               177.7      &    \\
    192          &  1999        &                28.8      &    \\
    196          &  1999        &                73.2      &    \\
    200          &  1999        &                74.2      &    \\
    202          &  1999        &                36.5      &    \\
    202--206     &  2000        &                83.1      &    \\
    206--209     &  2000        &               132.4      &    \\
    \hline
  \end{tabular}
  \caption{\label{t:lumi}
    Overview of the analysed integrated data luminosities.
    }
  \end{center}
\end{table}
                                
\begin{table}
  \small
  \begin{center}
    \begin{tabular}{|ll@{\hspace*{3em}}c|}\hline
      \multicolumn{2}{|l}{\LEP~1:\quad $\Zzero\to\ee,\mm$}              &\\ 
      \hline 
      0. & Preselection                           & see text\\ 
      1. & Modified acoplanarity                  & $0.11~\mathrm{rad} < \alpha < 2.0~\mathrm{rad}$\\ 
      2. & Polar angle of missing momentum vector & $|\cos\tpmiss| < 0.98 \quad $for $\pmiss > 2~\GeV $\\ 
      3. & Isolation of lepton tracks             &
                                $\max(\alpha_{\mathrm{iso}_1},\alpha_{\mathrm{iso}_2}) > 20^{\circ}$\\ 
      & &                       $\min(\alpha_{\mathrm{iso}_1},\alpha_{\mathrm{iso}_2}) > 10^{\circ}$\\  
      4. & Invariant mass of the lepton pair      & $20~\GeV < m_{\ell\ell} < 100~\GeV $\\ 
      5. & Photon and Conversion veto             & see text\\
      \hline
      \multicolumn{3}{c}{ }\\
      \hline
      \multicolumn{2}{|l}{\LEP~2:\quad $\Zzero\to\ee,\mm$}              &\\
      \hline 
      0. & Preselection                           & see text\\ 
      1. & Acoplanarity                           & $\pa >$ 0.15--0.20~rad\\
      2. & Polar angle of missing momentum vector & $|\cos\tpmiss| < 0.95 \quad $for $\pmiss > 5~\GeV $\\ 
      3. & Isolation of lepton tracks             &
                                $\max(\alpha_{\mathrm{iso}_1},\alpha_{\mathrm{iso}_2}) > 15^{\circ}$\\ 
      & &                       $\min(\alpha_{\mathrm{iso}_1},\alpha_{\mathrm{iso}_2}) > 10^{\circ}$\\  
      4. & Invariant mass of the lepton pair      &
                                                   $|m_{\ee}-m_{\Zzero}| <  8~\GeV $\\
         &                                        & $|m_{\mm}-m_{\Zzero}| < 10~\GeV $ \\ 
      5. & Photon veto                            & see text\\
      6. & Momentum in \textit{z}-direction       & $|p^z_1+p^z_2| < $ 50~\GeV\\
      \hline
      \multicolumn{3}{c}{ }\\
      \hline
      \multicolumn{2}{|l}{\LEP~1:\quad $\Zzero\to\nn$}              &\\
      \hline 
      1. & Cosmic muon and beam halo veto         & see text            \\
      2. & Number of identified electron tracks     & $N_e = $ 0 or 2    \\
      3. & Visible energy in electromagnetic calorimeter
         & $E_{\mathrm{Ecal}} >  12~\GeV, < 60~\GeV$                     \\
      4. & Transverse momentum of event           & $p_T > 6~\GeV$       \\
      5. & Direction of energy vector             & $|\cos\theta_{\vec{E}}| < 0.65$ \\
      6. & Energy in forward detector             & $E_{\mathrm{Fdet}} < 2~\GeV$      \\
     \multicolumn{3}{|l|}{Additional cuts for events with two electron tracks}   \\
      7. & Angle between tracks                   & $\Delta\phi< 10^\circ$       \\
      8. & Transverse momentum of event           & $p_T > 7~\GeV$        \\
      9. & Unassociated clusters in electromagnetic calorimeter
                                                  & $N_{\mathrm{unass.}} = 0$     \\
      \hline
    \end{tabular}
    \caption{\label{t:cuts_lep2}\label{t:cuts_lep1}
      A summary of the selection criteria.
      }
  \end{center}
\end{table}

\begin{table}
  \begin{center}
    \small
    \begin{tabular}{|c|r|r||r|r|r||c|}
      \hline
      \multicolumn{7}{|c|}{$\sqrts = 91.2~\GeV$}\\
      \hline
      \rb{Cut} &  
      \rb{Data}             & 
      \multicolumn{1}{c||}{\raisebox{-1ex}{Total}}&
      \rb{2-fermion}        & 
      \rb{4-fermion}        &  
      \rb{2-photon}         &  
      \multicolumn{1}{c|}{\rb{Signal}} \\
      & & \multicolumn{1}{c||}{\raisebox{1ex}{bkg.}} & & & &  \multicolumn{1}{c|}{\raisebox{1mm}{($m_{\Szero}$=30~GeV)}} \\ 
      \hline\hline
      \multicolumn{7}{|c|}{Electron channel}\\
      \hline
Preselection   
          &  122431   &129115   &128490       & 586.3     & 38.9      &  46.2\,\%       \\
$\alpha$  
          &    1560   &  1694   &  1628       &  58       &  8        &  33.4\,\%       \\
$|\cos\theta_{\pmiss}|$
          &    1500   &  1628   &  1571       &  55       &  2        &  32.8\,\%       \\
Lepton isolation
          &    1368   &  1466   &  1414       &  50       &  2        &  28.6\,\%       \\
$M_{\ell\ell}$
          &    1362   &  1462   &  1410       &  50       &  2        &  28.6\,\%       \\
Photon+Conversion veto
          &      45   &    55.2 &    20.5     &  34.4     &  0.3      &  28.6\,\%       \\
      \hline\hline
      \multicolumn{7}{|c|}{Muon channel}\\
      \hline
Preselection   
          &  109552   &115001   &114475       &  459.1      &  66.6     &  54.0\,\%     \\
$\alpha$  
          &    1575   &  1601   &  1526       &   58        &  17       &  40.2\,\%     \\
$|\cos\theta_{\pmiss}|$
          &    1549   &  1575   &  1512       &   57        &   6       &  40.0\,\%     \\
Lepton isolation
          &    1403   &  1470   &  1412       &   52        &   6       &  37.4\,\%     \\
$M_{\ell\ell}$
          &    1397   &  1467   &  1410       &   51        &   6       &  37.4\,\%     \\
Photon+Conversion veto
          &      66   &    53.6 &    17.0     &   35.4      &  1.2      &  35.0\,\%     \\
      \hline
    \end{tabular}
  \end{center}


  \begin{center}
    \begin{tabular}{|c|r|r||r|r|r||c|}
    \multicolumn{7}{c}{}\\
      \multicolumn{7}{c}{}\\
      \hline
      \rb{Cut} &  
      \rb{Data}             & 
      \multicolumn{1}{c||}{\raisebox{-1ex}{Total}}&
      \rb{$\nu\bar{\nu}\gamma$} &
      \rb{leptons}&
      \rb{other}  &  
      \multicolumn{1}{c|}{\rb{Signal}} \\
      & & \multicolumn{1}{c||}{\raisebox{1ex}{bkg.}} & & & & \\ 
      \hline\hline
      \multicolumn{7}{|c|}{Missing energy channel}\\
      \hline
      \multicolumn{6}{|c}{Events with 0 tracks} & $m_{\Szero}$=5~GeV, $\Szero\to\gamma\gamma$\\
      \hline
Preselection
          &    73     &  68.5   &  63.5       &     4.8   &     0.3   &  44.2\,\%       \\
$|\cos\theta_{\vec{E}}| < 0.65$
          &    54     &  51.1   &  48.1       &     2.7   &     0.3   &  38.4\,\%       \\
$E_{\mathrm{Ecal}} > 12~\GeV$
          &    14     &  10.7   &   9.8       &     0.6   &     0.3   &  30.0\,\%       \\
      \hline
      \multicolumn{6}{|c}{Events with 2 tracks} & $m_{\Szero}$=100~MeV, $\Szero\to\ee$\\
      \hline
Preselection
          &    30     &  21.6   &  4.5        & 11.6      &  5.5      &  29.2\,\%       \\
$|\cos\theta_{\vec{E}}| < 0.65$
          &    17     &  14.4   &  3.7        &  9.5      &  1.2      &  25.7\,\%       \\
$\Delta\phi < \degree{10}$  
          &    13     &   7.9   &  3.6        &  3.7      &  0.6      &  25.7\,\%       \\
$m_{\ee} < 2~\GeV$  
          &    12     &   7.9   &  3.6        &  3.7      &  0.6      &  25.7\,\%       \\
Charge $q_1\cdot q_2 = -1$
          &    10     &   6.3   &  3.6        &  2.1      &  0.6      &  25.2\,\%       \\
$N_{unass.} = 0$
          &     4     &   4.0   &  3.4        &  0.0      &  0.6      &  23.7\,\%       \\
$p_T > 7~\GeV$
          &     3     &   2.5   &  2.5        &  0.0      &  0.0      &  20.9\,\%       \\
$E_{\mathrm{Ecal}} > 12~\GeV$
          &     1     &   0.5   &  0.5        &  0.0      &  0.0      &  14.8\,\%       \\
      \hline 
  \end{tabular}
\caption{\label{t:cutflow_LEP1}
  Cutflow tables for the \LEP~1 analysis: Number of selected events
  after each cut. As an example the efficiencies for the signal
  process with $\Szero \to \bbar$ are given for the lepton channels,
  and with $\Szero \to \gamma\gamma$ and $\Szero \to \ee$ for the
  missing energy channel. The preselection in the missing energy
  channel includes the cuts 1, 4, 6 from the $\Zzero \to \nn$ channel
  in Table~\ref{t:cuts_lep1}, $E_{\mathrm{total}} > 4~\GeV$ and
  $|\cos\theta_i| < 0.9$ (see Section~\ref{s:misse_event_selection}).
  }
  \end{center}
\end{table}
\begin{table}
  \begin{center}
    \small
    \begin{tabular}{|c|r|r||r|r|r||c|}
      \multicolumn{7}{c}{}\\
      \hline
      \multicolumn{7}{|c|}{$\sqrts = 183$--$209~\GeV$}\\
      \hline
      \rb{Cut} &  
      \rb{Data}             & 
      \multicolumn{1}{c||}{\raisebox{-1ex}{Total}}&
      \rb{2-fermion}        & 
      \rb{4-fermion}        &  
      \rb{2-photon}         &  
      \multicolumn{1}{c|}{\rb{Signal}} \\
      & & \multicolumn{1}{c||}{\raisebox{1ex}{bkg.}} & & & &  \multicolumn{1}{c|}{\raisebox{1mm}{($m_{\Szero}$=90~GeV)}} \\ 
      \hline\hline
      \multicolumn{7}{|c|}{Electron channel}\\
      \hline

Preselection
          &    27708  &28183.5  &27720.0      &378.0      &85.5       &  49.1\,\%   \\
Lepton isolation
          &    24176  &24803.9  &24410.6      &314.3      &79.0       &  42.1\,\%       \\
$M_{\ell\ell}$
          &    708    &639.1    &547.9        &73.0       &18.2       &  37.7\,\%       \\
Photon-veto
          &    470    &477.1    &393.8        &67.9       &15.4       &  37.7\,\%       \\
$|\cos\theta_{\pmiss}|$
          &    118    &106.3    &57.4         &45.7       &3.2        &  34.8\,\%       \\
Acoplanarity
          &    67     &63.1     &25.4         &37.2       &0.5        &  28.7\,\%       \\
$|p^z_1+p^z_2|$
          &    54     &46.9     &12.8         &33.7       &0.4        &  28.7\,\%       \\ 
      \hline\hline
      \multicolumn{7}{|c|}{Muon channel}\\
      \hline

Preselection
          &    3042   &3115.6   &2818.8       &212.2      &84.6       &   64.7\,\%      \\
Lepton isolation
          &    2866   &2948.5   &2669.5       &195.9      &83.1       &   55.7\,\%      \\
$M_{\ell\ell}$
          &    803    &842.4    &733.3        &88.5       &20.7       &   49.3\,\%      \\
Photon-veto
          &    575    &629.3    &532.0        &80.9       &16.4       &   49.3\,\%      \\
$|\cos\theta_{\pmiss}|$
          &    111    &101.5    &45.8         &52.3       &3.4        &   45.5\,\%      \\
Acoplanarity
          &    66     &72.0     &26.7         &44.3       &1.0        &   37.5\,\%      \\
$|p^z_1+p^z_2|$
          &    43     &51.6     &12.2         &38.6       &0.8        &   37.5\,\%      \\
      \hline
    \end{tabular}
    \caption{\label{t:cutflow_LEP2}
      Cutflow tables for the \LEP~2 analysis: Number of selected
      events after each cut. As an example the efficiencies for the
      signal process $\Szero\Zzero\to\bbar\,\ell^+\ell^-$ are also
      given. The efficiencies are the average of the values at
      183--209~\GeV.  
      }
  \end{center}
\end{table}


\begin{table}
  \centering
  \small
\begin{tabular}{|c|c|r||c|c|c||r|}\hline
\rb{$\sqrt{s}$~(GeV)} &  
\rb{Data}             & 
\multicolumn{1}{c||}{\raisebox{-1ex}{Total}}&
\rb{2-fermion}        & 
\rb{4-fermion}        &  
\rb{2-photon}         &  
\multicolumn{1}{c|}{\rb{Signal}} \\
& & \multicolumn{1}{c||}{\raisebox{1ex}{bkg.}} & & & &  \multicolumn{1}{c|}{\raisebox{1mm}{($m_{\Szero}$=30~GeV)}} \\ 
\hline\hline
\multicolumn{1}{|c}{ }& \multicolumn{6}{c|}{Electron channel}\\
\hline
91.2 &  45 & 55.2\erro{3.0}{3.0} & 20.5 & 34.4 & 0.3  & 15.61\erro{0.31}{0.47}\\ \hline
183  &   7 &  3.6\erro{0.1}{0.3} & 1.4  & 2.1  & 0.1  &  0.91\erro{0.02}{0.03}\\
189  &  18 & 13.7\erro{0.4}{1.0} & 4.2  & 9.5  & 0.0  &  2.42\erro{0.04}{0.09}\\
192  &   0 &  2.2\erro{0.1}{0.2} & 0.7  & 1.5  & 0.0  &  0.37\erro{0.01}{0.01}\\
196  &   6 &  5.7\erro{0.2}{0.4} & 2.0  & 3.7  & 0.0  &  0.87\erro{0.01}{0.03}\\
200  &   4 &  4.8\erro{0.2}{0.3} & 1.2  & 3.5  & 0.1  &  0.81\erro{0.01}{0.03}\\
202  &   5 &  2.5\erro{0.1}{0.2} & 0.6  & 1.9  & 0.0  &  0.39\erro{0.01}{0.01}\\
202--206
     &   5 &  5.0\erro{0.2}{0.4} & 0.7  & 4.2  & 0.1  &  0.86\erro{0.01}{0.03}\\
206--209
     &   9 &  9.4\erro{0.3}{0.7} & 2.0  & 7.3  & 0.1  &  1.34\erro{0.02}{0.05}\\ \hline
$\sum(\ge$ 183)
     &  54 & 46.9\erro{0.6}{3.5} & 12.8 & 33.7 & 0.4  &  7.97\erro{0.06}{0.25} \\
\hline\hline
\multicolumn{1}{|c}{ }& \multicolumn{6}{c|}{Muon channel}\\
\hline
91.2 &  66 & 53.6\erro{2.7}{2.1} & 17.0 & 35.4 & 1.2  & 21.55\erro{0.45}{0.69}\\ \hline
183  &   5 &  4.4\erro{0.1}{0.2} & 1.6  & 2.7  & 0.1  &  1.20\erro{0.01}{0.05}\\
189  &   9 & 13.7\erro{0.1}{0.7} & 4.0  & 9.5  & 0.2  &  2.96\erro{0.03}{0.11}\\
192  &   2 &  2.5\erro{0.1}{0.1} & 0.6  & 1.9  & 0.0  &  0.46\erro{0.01}{0.02}\\
196  &   6 &  6.1\erro{0.1}{0.3} & 1.2  & 4.7  & 0.2  &  0.96\erro{0.01}{0.04}\\
200  &   5 &  5.7\erro{0.1}{0.3} & 1.3  & 4.3  & 0.1  &  0.89\erro{0.01}{0.03}\\
202  &   3 &  2.9\erro{0.1}{0.1} & 0.6  & 2.3  & 0.0  &  0.43\erro{0.01}{0.02}\\
202--206
     &   9 &  6.0\erro{0.1}{0.3} & 0.9  & 5.0  & 0.1  &  1.00\erro{0.01}{0.04}\\
206--209
     &   4 & 10.3\erro{0.1}{0.5} & 2.0  & 8.2  & 0.1  &  1.53\erro{0.02}{0.06}\\ \hline
$\sum(\ge$ 183)
     &  43 & 51.6\erro{0.3}{2.5} &12.2  & 38.6 & 0.8  &  9.43\erro{0.06}{0.37} \\
\hline
\multicolumn{7}{c}{\vspace*{1ex}}\\
\hline
\multicolumn{1}{|c}{ }& \multicolumn{6}{c|}{Missing energy channel}\\ \hline
\rb{$\sqrt{s}$~(GeV)} &
\rb{Data}             &
\multicolumn{1}{c||}{\raisebox{-1ex}{Total}}&
\rb{$\nu\bar{\nu}\gamma$} &
\rb{$n\cdot\gamma$}       &
\rb{other}                &
\multicolumn{1}{c|}{\rb{Signal}}\\
& & \multicolumn{1}{c||}{\raisebox{1ex}{bkg.}} & & & & \multicolumn{1}{c|}{\raisebox{1mm}{($m_{\Szero}$=30~GeV)}} \\ 
\hline
91.2 &  15 & 11.3\erro{1.1}{0.2} & 10.3 & 0.3  & 0.7  &175.07\erro{3.85}{2.80}\\
\hline
\end{tabular}

\caption{\label{t:d_b_error} 
  Summary of selected data events,
  background Monte Carlo and signal expectation for a 30~GeV \SM
  Higgs boson in the decay-mode independent searches. The first
  error is statistical and the second error is systematic. 
  }
\end{table}


\begin{table}
\begin{center}
\small
\begin{tabular}{|l|p{7ex}|p{7ex}|p{7ex}|p{7ex}|p{7ex}|}\hline
\multicolumn{5}{|c|}{Electron channel -- uncertainties in \% }\\
\hline
        & \multicolumn{2}{c|}{91~GeV}
        & \multicolumn{2}{c|}{183--209~GeV} \\
\cline{2-5}
\raisebox{1.3ex}[0ex][0ex]{Source}
                &  Bkg.  &  Sig.  &  Bkg.  &  Sig.   \\ 
\hline                                              
Electron-ID     &  0.8   &  0.8   &  1.3   &  1.3    \\
Energy          &  ---   &  ---   &  1.2   &  1.5    \\
Isolation angle &  1.3   &  2.6   &  4.3   &  2.8    \\
Trk. resolution &  2.3   &  1.3   &  2.2   &  1.3    \\
ISR/FSR         &  2.4   &  ---   &  4.7   &  ---    \\
$\alpha_{\mathrm{em}}$                              
                &  4.0   &  ---   &  0.4   &  ---    \\
Luminosity      &  0.5   &        &  0.2   &  ---    \\
\hline                                              
Total systematics& 5.4   &  3.0   &  7.0   &  3.7    \\
\hline\hline                                        
Statistics      &  5.5   &  2.0   &  3.1   &  0.9    \\
\hline
\end{tabular}
\bigskip\bigskip

\begin{tabular}{|l|p{7ex}|p{7ex}|p{7ex}|p{7ex}|p{7ex}|}\hline
\multicolumn{5}{|c|}{Muon channel -- uncertainties in \% }\\
\hline
        & \multicolumn{2}{c|}{91~GeV}
        & \multicolumn{2}{c|}{183--209~GeV}\\
\cline{2-5}
\raisebox{1.3ex}[0ex][0ex]{Source}
                &  Bkg.  &  Sig.  &  Bkg.  &  Sig.   \\ 
\hline                                             
Muon-ID         &  1.5   &  1.5   &  2.8   &  2.8    \\
Momentum        &  ---   &  ---   &  1.9   &  1.3    \\
Isolation angle &  0.2   &  2.1   &  1.7   &  2.0    \\
Trk. resolution &  2.7   &  1.9   &  2.2   &  1.1    \\
ISR/FSR         &  2.0   &  ---   &  2.3   &  ---    \\
$\alpha_{\mathrm{em}}$                             
                &  1.2   &  ---   &  0.2   &  ---    \\
Luminosity      &  0.5   &  ---   &  0.2   &  ---    \\
\hline                                             
Total systematics& 3.9   &  3.2   &  5.0   &  3.8    \\
\hline\hline                                       
Statistics      &  5.1   &  2.1   &  1.7   &  1.0    \\
\hline
\end{tabular}
\caption{
  Systematic uncertainties in percent for background and signal. For
  $\sqrt{s} = 91~\GeV$ the uncertainties are given for $\mS =
  30~\GeV$, for $\sqrt{s} > 91~\GeV$ they are shown for $\mS =
  60~\GeV$.
  }
\label{tab:sys_err}
\end{center}
\end{table}

\begin{table}
\centering
{\small
\begin{tabular}{|p{15.5ex}|p{6ex}|p{6ex}|p{6ex}|p{6ex}|p{6ex}|p{6ex}|p{6ex}|p{6ex}|p{6ex}|p{7ex}|}
\hline
&\multicolumn{10}{c|}{\vnr{2.5ex}Measurement of $\tilde{K}$ in bins of
  10~\GeV width} \\
\hline
Bin  & 1     & 2      & 3      & 4      & 5      & 6      & 7      & 8      & 9      & 10      \\
\hline
Mass  (GeV)
     &0--10  &10--20  & 20--30 & 30--40 & 40--50 & 50--60 & 60--70 & 70--80 & 80--90 & 90--100 \\
\hline
\hline
$\vnr{2.5ex}\tilde{K}$\hfill$\times 10^3 \quad(\GeV^{-1})$ 
     & $2.1$ & $-2.4$ & $-4.9$ & $-2.8$ & $-7.1$ & $5.8$  & $-33.5$ & $-45.2$ & $-18.6$ & $200.2$    \\
\hline
$\vnr{2.5ex}\Delta(\tilde{K})_{\mathrm{stat.}}$\hfill$\times 10^3 \quad(\GeV^{-1})$ 
     & $2.9$ & $4.4$   & $4.8$  & $6.4$  & $14.3$  & $23.7$  & $21.7$  & $30.5$  & $66.9$  & $166.4$   \\
$\vnr{2.5ex}\Delta(\tilde{K})_{\mathrm{sys.}}$\hfill$\times 10^3 \quad(\GeV^{-1})$ 
     & $0.9$ & $0.9$  & $0.6$  & $0.5$  & $1.3$  & $3.3$  & $4.5$  & $7.0$  & $16.5$  & $37.4$   \\
\hline
\end{tabular}
}
\caption{\label{t:bin_klim}
  Bin-wise measurement of $\tilde{K}$ for the mass range
  0--100~\GeV with $\Delta m = 10~\GeV$. To fit a theoretical
  distribution $\tilde{K}$ to these values, the
  correction matrix $\hat{C}$ from Table~\ref{t:correction_matrix}
  must be applied first.  
  }
\end{table}
\begin{table}
  \begin{center}
    \begin{displaymath}
      \hat{C} = 
      \left(
        \begin{array}{cccccccccc}
         0.33&0.04&0.02&0.01&0.00&0.00&0.00&0.00&0.00&0.00\\
         0.41&0.53&0.02&0.01&0.00&0.00&0.00&0.00&0.00&0.00\\
         0.17&0.27&0.11&0.04&0.01&0.00&0.00&0.00&0.00&0.00\\
         0.09&0.11&0.14&0.17&0.05&0.01&0.00&0.00&0.00&0.00\\
         0.00&0.05&0.17&0.21&0.28&0.09&0.01&0.00&0.00&0.00\\
         0.00&0.00&0.16&0.20&0.27&0.34&0.11&0.01&0.00&0.00\\
         0.00&0.00&0.11&0.14&0.16&0.29&0.37&0.09&0.01&0.00\\
         0.00&0.00&0.10&0.10&0.10&0.15&0.25&0.43&0.06&0.00\\
         0.00&0.00&0.09&0.07&0.07&0.08&0.11&0.27&0.46&0.03\\
         0.00&0.00&0.09&0.07&0.05&0.04&0.06&0.11&0.32&0.32\\
        \end{array}
      \right)
    \end{displaymath}
    \caption{
      Correction matrix for mass resolution. For a given theory to be
      tested with a distribution of $\tilde{K}$ values in the 10 mass
      bins, $\vec{\kappa} =
      \left(\tilde{K}_{1},\ldots,\tilde{K}_{10}\right)$, the vector
      $\vec{\kappa}$ has to be multiplied by the matrix $\hat{C}$ to
      account for mass resolution effects. The corrected vector
      $\vec{\kappa}' = \hat{C}\vec{\kappa}$ can then be fitted to the
      measured values of $\tilde{K}$ from Table~\ref{t:bin_klim}.  
      }
    \label{t:correction_matrix}
  \end{center}
\end{table}

\clearpage

\clearpage

\clearpage

\clearpage

\clearpage

\clearpage

\clearpage

\clearpage
\begin{figure}
  \centering
  \includegraphics[width=9cm]{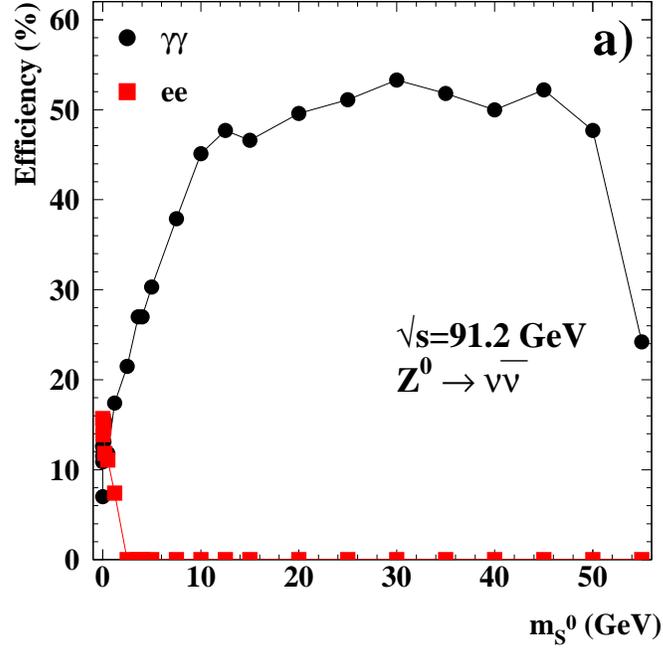}\\[4ex]
  \includegraphics[width=9cm]{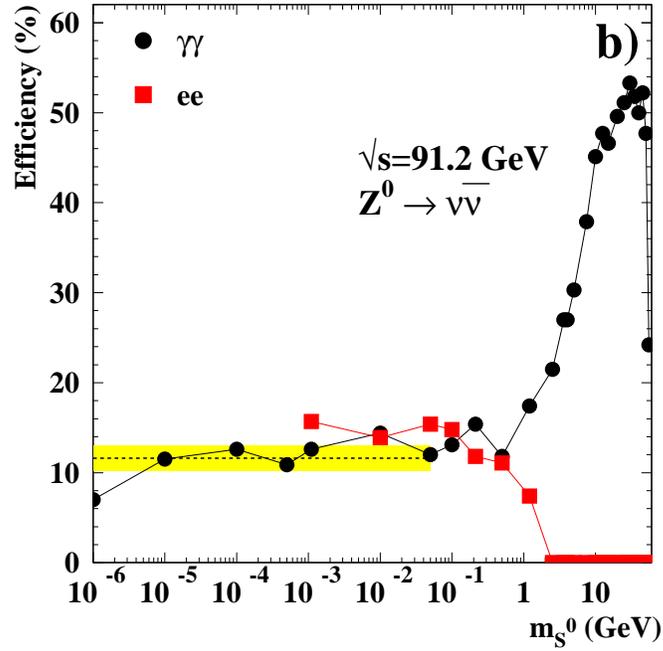}
  \caption{\label{f:lep1_nneff}
    The efficiency versus the \Szero mass at $\sqrt{s}=91$~GeV for the
    decay $\Szero\Zzero\to\gamma\gamma\nn$ and
    $\Szero\Zzero\to\ee\nn$. a)~in linear mass scale and b)~in
    logarithmic mass scale.
    The dashed line indicates the average of the efficiencies
    for $\mS \le 50~\MeV$. 
    The shaded bands show the typical error on the efficiencies in
    this region.
  }
\end{figure}

\newpage

\newpage

\newpage

\newpage

%
%
\end{document}